\documentclass[12pt,preprint]{aastex}		

\usepackage{natbib}
\usepackage{graphics}
\usepackage{psfig}
\usepackage{latexsym,amssymb}

\lefthead{Bianchi et al}
\righthead{Classification of UV sources}

\newcommand{\fngr}{[FUV-NUV]~{\it vs}~[g-r]~}

\newcommand{\fnnr}{[FUV-NUV]~{\it vs}~[NUV-r]~}
\newcommand{\nggi}{[NUV-g]~{\it vs}~[g-i]~}
\newcommand{\fggi}{[FUV-g]~{\it vs}~[g-i]~}
\newcommand{\nggr}{[NUV-g]~{\it vs}~[g-r]~}
\newcommand{\fggr}{[FUV-g]~{\it vs}~[g-r]~}

\newcommand{\Teff}{\mbox{$T_{\rm eff}$}}

\newcommand{\Rsun}{\mbox{$R_{\odot}$}}

\newcommand{\ebv}{$E(B-V)$}

\newcommand{\fuvmag}{\mbox{\it FUV~}}
\newcommand{\nuvmag}{\mbox{\it NUV~}}
\slugcomment{Submitted for publication in the Special GALEX ApJS Issue}

\begin{document}

\title{Statistical Properties of the GALEX/SDSS matched source catalogs, 
and classification of the UV sources.}

\author{Luciana Bianchi\altaffilmark{1}, 
Lino Rodriguez-Merino\altaffilmark{1},
Maurice Viton\altaffilmark{2}, 
Michel Laget\altaffilmark{2},
Boryana Efremova\altaffilmark{1}, 
James Herald\altaffilmark{1},
Alberto Conti\altaffilmark{3},Bernie Shiao\altaffilmark{3}, 
Armando Gil de Paz\altaffilmark{4},
Samir Salim\altaffilmark{5}, 
A. Thakar\altaffilmark{1}, 
Peter G. Friedman\altaffilmark{6},
S.C. Rey\altaffilmark{6}, 
David Thilker\altaffilmark{1},
Tom A. Barlow\altaffilmark{6}, 
Tamas Budavari\altaffilmark{1}, Jose Donas\altaffilmark{2},
Karl Forster\altaffilmark{6},
Timothy M. Heckman\altaffilmark{1},
Young-Wook Lee\altaffilmark{8},
Barry F. Madore\altaffilmark{9},
D. Christopher Martin\altaffilmark{6},
Bruno Milliard\altaffilmark{2},
Patrick Morrissey\altaffilmark{6},
Susan G. Neff\altaffilmark{7},
R. Michael Rich\altaffilmark{5},
David Schiminovich\altaffilmark{6},
Mark Seibert\altaffilmark{6},
Todd Small\altaffilmark{6},
Alex S. Szalay\altaffilmark{1},
Ted K. Wyder\altaffilmark{6},
Barry Y. Welsh\altaffilmark{10}, 
Sukyoung K. Yi \altaffilmark{8}
 }
\altaffiltext{1}{Dept. of Physics and Astronomy,
The Johns Hopkins University, 
3400 N. Charles St., Baltimore,
MD 21218, USA. Email: bianchi@pha.jhu.edu} 
\altaffiltext{2}{Laboratoire d'Astrophysique de Marseille, BP 8, Traverse
du Siphon, 13376 Marseille Cedex 12, France}
\altaffiltext{3}{Space Telescope Science Institute, Baltimore, MD 21218 }
\altaffiltext{4}{Univ. Complutense, Madrid, Spain}
\altaffiltext{5}{Department of Physics and Astronomy, University of
California, Los Angeles, CA 90095}
\altaffiltext{6}{California Institute of Technology, MC 405-47, 
1200 East California Boulevard, Pasadena, CA 91125h}
 \altaffiltext{7}{Laboratory for Astronomy and Solar Physics, NASA Goddard
Space Flight Center, Greenbelt, MD 20771}
\altaffiltext{8}{Center for Space Astrophysics, Yonsei University, Seoul
120-749, Korea}
\altaffiltext{9}{Observatories of the Carnegie Institution of Washington,
813 Santa Barbara St., Pasadena, CA 91101}
\altaffiltext{10}{Space Sciences Laboratory, University of California at
Berkeley, 601 Campbell Hall, Berkeley, CA 94720}

\begin{abstract}
We use the {\it Galaxy Evolution Explorer} (GALEX)  
{\it Medium} and {\it All-Sky-Imaging Survey} (MIS \& AIS) data 
from the first public data release (GR1),
matched to the Sloan Digital Sky Survey (SDSS) DR3 catalog, 
to perform source classification. The GALEX surveys provide 
photometry in far- and  near-UV bands 
and the SDSS in five optical bands  ({\it u,g,r,i,z}).
The GR1/DR3 overlapping areas are 363[83]deg$^2$ for the GALEX AIS[MIS],
for sources within the 0.5~$^{\circ}$ central area of the  GALEX fields.
 Our sample covers mostly $|$b$|$$>$30~$^{\circ}$   galactic 
latitudes. 
We present statistical properties of the GALEX/SDSS matched sources  catalog,
containing $>$2~10$^6$ objects detected in at least one UV band.
We classify the matched sources by comparing the seven-band 
photometry  to  model colors   constructed for different classes
of astrophysical objects. For sources with 
photometric errors $<$0.3~mag, the corresponding 
typical AB-magnitude limits 
are   m$_{FUV}$$\sim$21.5,  m$_{NUV}$$\sim$22.5 for AIS, and 
m$_{FUV}$$\sim$24,  m$_{NUV}$$\sim$24.5 for MIS. 
At AIS depth, the number of Galactic  and 
extragalactic objects are comparable, but the latter 
predominate in the MIS. 
Based on  our stellar models, we estimate 
the GALEX surveys detect hot 
White Dwarfs  throughout the Milky Way halo (down to a 
 radius of 0.04\Rsun~ at MIS depth), 
providing an unprecedented improvement in  the 
Galactic WD census.
Their observed surface density is consistent with
 Milky Way model predictions. 
We also select low-redshift QSO candidates, extending the
known QSO samples to lower magnitudes, and providing candidates 
for  detailed {\it z}$\approx$1 follow-up investigations.
SDSS optical spectra available for a large  subsample 
confirm the classification for the photometrically selected candidates 
with   97\% purity for single hot stars, 
$\approx$45\%(AIS)/31\%(MIS) for 
binaries containing a hot star and a cooler companion,
and about 85\% for QSOs. 
\end{abstract}


\section{Introduction}
\label{s_intro}
Several recent and ongoing survey projects are 
providing a wealth of new  data that  
allow us to refine theoretical models, and our global understanding 
of stars, galaxies, and the
evolution of the Universe. The {\it Galaxy Evolution Explorer} (GALEX, 
Martin et al. 2005)
is performing the first survey of the sky in two broad bands:
the far ultraviolet (FUV) and the near ultraviolet (NUV). 
The {\it Sloan Digital Sky Survey} (SDSS) 
is scanning one fourth of the sky in five optical bands, {\it u, g, r, i}
and {\it z}. 
Bianchi et al. (2005) 
explored the potential of the multi-band photometric 
catalogs to classify astrophysical objects, using GALEX 
early release data 
and  showed that the UV bands are very sensitive 
to the detection of hot stellar objects and to the interstellar extinction by dust,
as well as to the identification of low-redshift QSOs, as expected. 
Star counts are 
important because they delineate galactic structure, and thus give clues to 
 galaxy formation and 
evolution. The GALEX UV surveys provide unbiased detection of 
evolved, hot post-AGB objects, and of binary stellar 
systems containing white dwarfs (WD), which will allow us  to  
populate the observational Hertzprung-Russel diagram
 of hot evolved objects. 
Late evolutionary phases are extremely important for the yield of processed
 elements  (e.g. Marigo 2004) 
and the consequent chemical evolution of the 
interstellar medium (ISM). However, the post-AGB phases  are 
 much less understood than the early evolutionary phases,
 due to the scarcity 
of known objects. Such evolved objects are  elusive in surveys at longer 
wavelengths  both because of their
low luminosity (at optical wavelengths)  
and high effective temperatures, to which optical colors
are insensitive, and because they are extremely short-lived. 
We use the GALEX deep sensitivity to significantly extend
the census of hot evolved objects in the Milky Way.

In this paper we analyze the characteristics of a database we obtained by 
matching the GALEX GR1 (MIS and AIS) and SDSS DR3 
(Abazajian et al. 2005)  imaging catalogs.
 Both the GALEX GR1, and the SDSS DR3 
data for the matched sources are available from the MAST archive. 
In section \ref{s_obs} we describe the characteristics of 
the GALEX and SDSS photometric data used in this work.
 In section \ref{s_match} we describe 
 our final  catalog of  GALEX/SDSS matched sources. 
In section \ref{s_analysis}  we use color-color diagrams 
 for classification of the sources. 
The results are discussed  in section \ref{s_disc}.

\section{Observations}
\label{s_obs}
This work makes use of  photometric data collected by two of the largest 
astronomical surveys  currently in progress, the GALEX mission 
and the SDSS project. 
Here  we recall the basic characteristics of the imaging data from the two 
instruments,  then we 
describe the matched source catalog constructed for this analysis. 

\subsection{GALEX data}
The GALEX mission 
(Martin et al. 2005) is performing a series of 
imaging and spectroscopic sky surveys in two 
 ultraviolet bands, 
FUV and NUV. The instrument consists of a 50 cm diameter modified Ritchey-Chr{\'e}tien telescope 
providing a very wide field of view 
(1.25$^{\circ}$ diameter) with good astrometric quality across most of 
the field, and a resolution 
of $\approx$ 4.5-6$''$ [FUV-NUV] (Morrissey et al. 2005).
 In this work we limit 
the analysis to the sources in the inner 0.5$^{\circ}$ radius of the GALEX field of view, for best 
astrometric and photometric quality.

The GALEX photometric data cover the wavelength range from 
1344 {\AA} to 2831 {\AA} with 
two broad bands, the FUV passband (1344 - 1786 \AA) 
with $\lambda$$_{eff}$ = 1528 {\AA} and 
the NUV band (1771 - 2831 \AA) with $\lambda$$_{eff}$ = 2271 {\AA}
(Morrissey et al. 2005). 
 Imaging surveys are 
carried out with different depth and coverage. 
In this work we use data from the All-sky Imaging Survey 
(AIS), that has typically $\sim$100 second  exposures with a 
5~$\sigma$ NUV limiting magnitude  of  
m$_{NUV}$$\sim$20.8, and from the Medium Imaging Survey (MIS),
with typical  $\sim$1500 second exposures and  
 limiting magnitude  of 
m$_{NUV}$$\sim$22.7 in the AB system (Morrissey et al. 2005). 
The photometric system used in this paper 
is based on the AB magnitude scale 
(Oke \& Gunn 1983), although in 
the color-color diagrams
we also provide the Vega-mag scale, 
to facilitate comparison with other work. 

We note that the GALEX flux calibration  has been revised since the 
early release data  (used e.g. by Bianchi et al.(2005)
in  similar work), and the current processing and calibration 
brings a total shift in the FUV-NUV color of $\sim$0.12\,mag,  
such that the GR1 photometry is 
giving redder colors than the early release data by about this amount. 
Zero points of 18.82~mag and 20.08~mag for the FUV and NUV magnitudes 
respectively (AB system) are used in the GALEX GR1 catalog. 

\subsection{Sloan Digital Sky Survey}
The SDSS project (York et al. 2000)
is mapping one fourth of the sky in five broad optical bands: {\it u g r i z}
(Fukugita et al. 1996), 
using a dedicated 2.5m telescope with a wide-field 
of view and a 0.5-meter telescope for  photometric calibration. 
An automated image-processing system measures photometric and astrometric properties of each 
source (Pier et al. 2003). 
The SDSS photometric system covers from 3000 {\AA} to 11000 {\AA} with five non-overlapping 
pass-bands. The {\it u} filter peaks at 3500 {\AA} with a width of 600 {\AA}, the {\it g} is 
a blue-green band centered at 4800 {\AA} with a width of 1400 {\AA}, {\it r} is a red band 
centered at 6250 {\AA} with a width of 1400 {\AA}, {\it i} is a far red filter centered at 
7700 {\AA} with a width of 1500 {\AA},
 and the near-infrared passband {\it z} is centered at 
9100 {\AA} with a width of 1200 {\AA}.

\section{The GALEX-SDSS Matched Source Catalog}
\label{s_match}
\begin{figure} 
\psfig{file=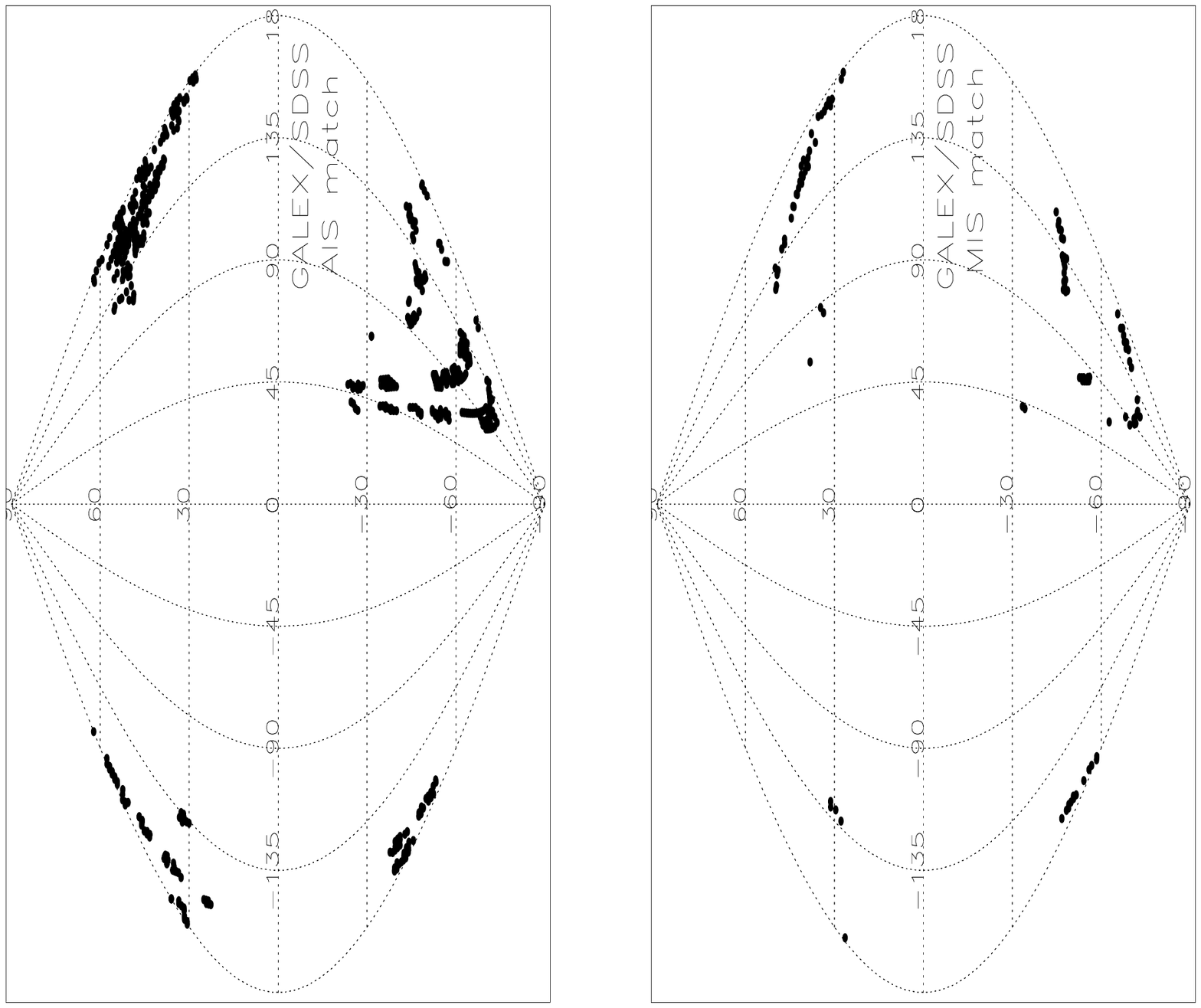,width=\hsize,angle=-90}
\caption{The location of the common areas between the GALEX GR1 release and the 
SDSS DR3 release are shown in galactic coordinates. 
Upper plot: the All-sky Imaging Survey carried out by GALEX has 622 fields matching the SDSS, 
the total unique area is 363$\pm$3 deg$^2$, when we restrict the GALEX fields to the central
1 degree diameter. Lower plot: The GALEX Medium Imaging Survey has 112 fields 
in common with SDSS DR3 covering a unique overlap area of 86$\pm$1 deg$^2$ .}
\label{plot_area_match}
\end{figure}

In the GALEX GR1 release there are 622  AIS fields and 112 MIS fields
which overlap areas of the sky observed by the SDSS. 
Figure \ref{plot_area_match} displays the GALEX/SDSS 
overlapping fields in galactic coordinates. 
As a starting point we used  a GALEX GR1/ SDSS DR3 matched catalog 
available from the MAST database. The match between GALEX and 
SDSS sources was done based on position using  a 4$''$ match 
radius (Budavari et al. 2004). 

\subsection{Calculation of areas of overlap}
\label{s_olap}
In order to calculate the density of sources from our classification work,
we 
determined the total areas of overlap between the 
GALEX GR1 and SDSS DR3 releases 
with the following  method. 
We use in our analysis  only sources within the central 
1$^{\circ}$ of the GALEX field, therefore for each GALEX  
field we considered an effective radius of 
0.5$^{\circ}$. Partial overlap among some GALEX fields is
 not removed in the on-line GALEX database, hence sources
observed in more than one field have
multiple entries in the catalog. 
We wrote a code that scans the entire sky 
and calculates the unique area covered 
by the GALEX fields included in our matched catalog, and then calculates 
the part of this area covered also by the SDSS. 
Three subsequent steps are performed: 
a) the celestial sphere is divided  in small (0.05 degrees on a side), 
approximately square,   elements, 
b) the distance between the center of each area element 
in the grid to the GALEX 
field centers is used to determine whether the element was included 
in our survey, and thus contributed to the total area, and 
c) a check is performed on each area element included 
in our GALEX coverage to assess whether  it 
was also included in the area observed by the SDSS.
The resulting unique-coverage areas of the matched catalogs are
 363 deg$^2$ for AIS and 86 deg$^2$ for MIS. 
These figures will be used to determine the density of sources 
(number/deg$^2$) in the next 
sections. Note that the total areas refer to a 0.5~degree radius GALEX field,
corresponding to our sample restriction.
 See also Figure 1 of Bianchi et al. (2006b). 
The error in our area determination was estimated by running several
 tests on a small 
area of the sky, including thousands of field position simulations. 
The maximum estimated uncertainty in the areas from our procedure is 
3 deg$^2$ and 1 deg$^2$ for AIS and MIS respectively. 

We note that the SDSS database sky partition 
(hierarchical triangular mesh, HTM) 
could not be used to estimate the area coverage 
due to the HTM's  intrinsic inability to return accurate areas, and to 
an error in the areas and vertices of the triangular tassels, which
prevented calculation of
exact areas of the HTM tassels.

\subsection{The final catalog} 
\label{s_unique}
We used as a starting point the matched catalogs  available from the MAST 
archive, which include all sources from the GALEX GR1 MIS and AIS,  
and from the SDSS DR3 releases, 
matched according to their position using a match radius of 4$''$.
The matched database contains a total of over 2 Million objects
 (1.4M from AIS and 0.9M from MIS) detected in at least one UV band. 
In order  to proceed to the analysis of the sources, we first eliminated 
multiple entries for the same source.

GALEX sources observed in more than one pointing result in 
multiple entries for the same source in the on-line catalog. 
Eliminating GALEX duplicate entries for the same source 
reduced the number of sources by 23\% for AIS and by 15\% for MIS.
In addition, some  GALEX 
sources have more than one SDSS counterpart due to the different 
spatial resolution: GALEX 
has a point spread function of $\sim$4.5-6$"$  while the  SDSS' psf 
is $\sim$1.4$"$. For these sources  
the analysis of photometric colors would be meaningless, 
since the GALEX photometry would be a 
composite measurement  of different objects. 
Therefore, GALEX sources with multiple matches 
  must be excluded from the classification analysis,
but their fraction must be counted for source statistics. 
Removal of GALEX sources with multiple optical counterparts 
within the match radius reduced the number of sources 
by 13\% for AIS and by 11\% for MIS. 
Our object density counts (section 5) take into account the 
 fraction of sources eliminated because having more than one 
optical counterpart.    After 
eliminating duplicate GALEX detections, and UV sources with 
multiple SDSS matches,  our final catalog
 contains 1,074,460 AIS  sources and 752,190 MIS sources.

Finally, we 
must evaluate 
the possible contamination by spurious matches, and whether the purity of the
sample would be improved by decreasing the match radius.
In this work we considered all sources with a match radius up to 4$''$
(but excluded multiple matches, as explained above). 
%
%
 When we apply error cuts of 0.15/0.1/0.1\,mag (FUV/NUV/optical) to the 
sample,
the distance between GALEX coordinates and SDSS coordinates is
less than 2.5$''$ for 99/96\% of the sources [AIS/MIS]. 
If  no error cuts are applied, a
higher number of matched sources with GALEX and SDSS coordinates
differing by more than 2.5$''$ is found over the entire catalog. 
As a result of the photometric error cuts 
applied in our analysis,  possible spurious matches
 are not expected to  significantly affect  the results. 

In fact, the UV sky is rather ``empty'', i.e. much more scarcely populated 
than the sky at optical wavelengths, due to the paucity of hot
stars relative to the vast majority of cooler objects. Therefore, a
possible spurious match (positional coincidence) would probably 
occur if multiple optical sources are detected around the 
position of the UV source. Such cases are eliminated as
explained above. Because we use both FUV and NUV bands, and our optical
photometry includes the {\it u} and {\it g} bands, we 
expect that a hot object appearing as both an FUV and NUV source would
 be detected in  {\it u} and {\it g}.  Random positional
associations would be much more likely to  occur
if we considered matches of UV sources with red or 
IR bands. We did not impose in our final analysis samples 
 that good photometric measurements exist for the UV sources
also in  {\it i} and {\it z} bands. Such requirement would 
considerably limit the final sample of hot stellar objects,
in addition to possibly favouring spurious matches. 
A purity of over 95\% for matches of UV sources with optical sources
was predicted by simulations conducted during the design phases.  

We also checked for possible artifacts
in the matched catalog. Instrumental artifacts would likley
be eliminated by the matching process itself, 
however a caveat remains
about possible contamination of our samples by SDSS sources
that the pipeline generates by
``shredding'' galaxies, i.e. deblending extended objects into a number of 
individual photometric sources. We have investigated the problem and 
we believe such sources
 cannot be reliably eliminated with  the use of the
current SDSS or GALEX photometric flags. 
Therefore we specifically searched whether
any of our selected sources (section 5) falls within the radius
of any known galaxy, using a list of galaxies with 
R$_{25}$$>$1arcmin extracted 
from the  catalog of de Vaucouleurs et al. (1991).
We found only one case in our AIS catalogs of WD, and 3 cases in
the MIS catalog, of point-like sources which are in fact portions
of large galaxies.  A few sources out of several thousand
QSO candidates were also identified as ``shredded galaxies''.

\subsection{Properties of the matched GALEX/SDSS source catalog. 
Analysis sample definition.}
\label{s_sample}

\begin{figure}[!ht]
\centerline{
\psfig{file=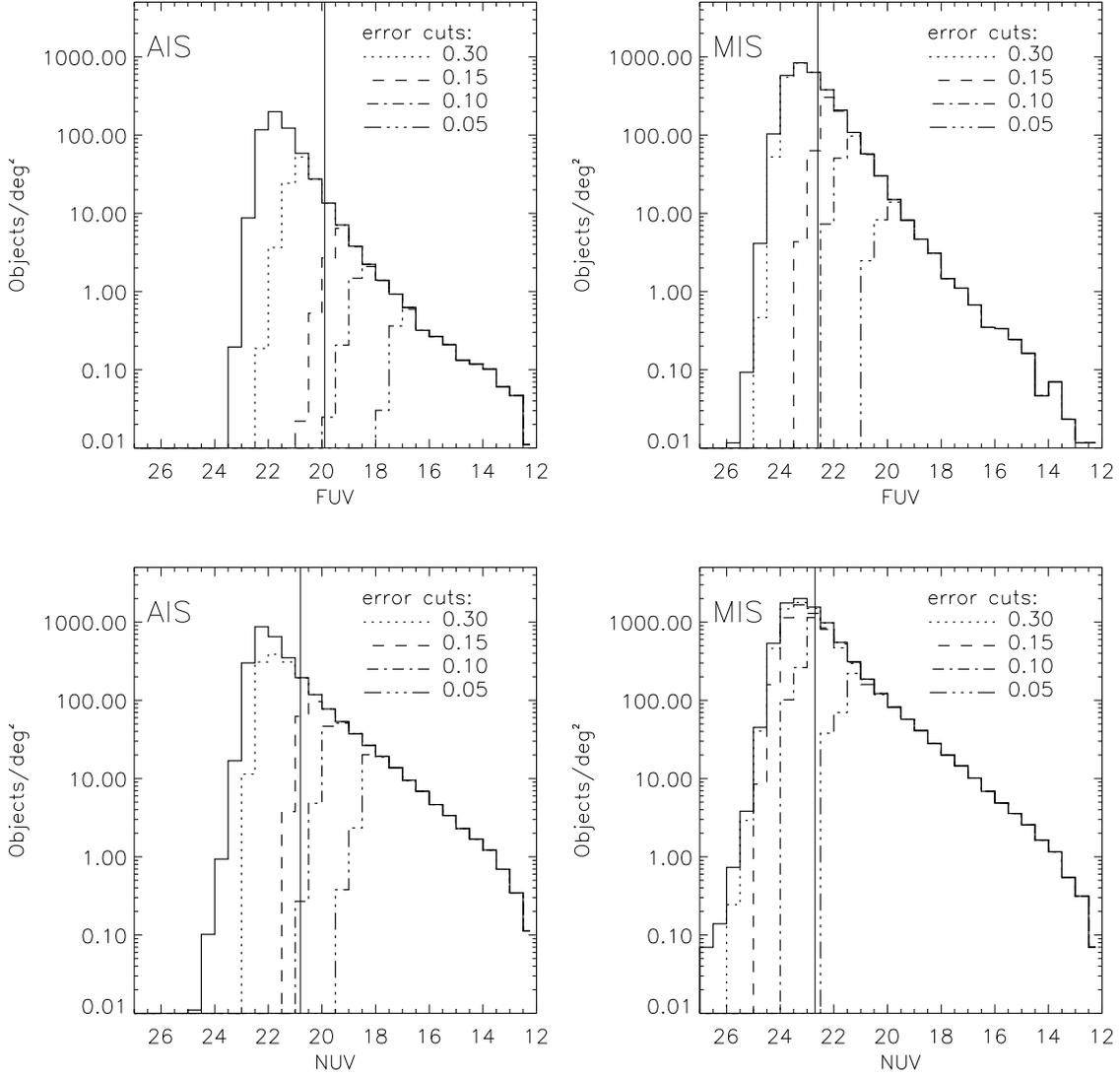,width=6.in,clip=}
}
\caption{The distribution 
of sources from our GALEX/SDSS matched catalog as a function of UV 
magnitudes (FUV in the upper panels, NUV in the lower panels), 
separately for AIS and MIS. 
The solid line is the total number of sources (after duplications
 have been removed as explained 
in the text), other lines show the histograms when error cuts are
 applied. Vertical lines display
the 5$\sigma$ detection limit  for a typical exposure 
(from the GALEX GR1 documentation via the MAST  web site).}
\label{plot_hist1}
\end{figure}

\begin{figure}[!ht]
\centerline{
\psfig{file=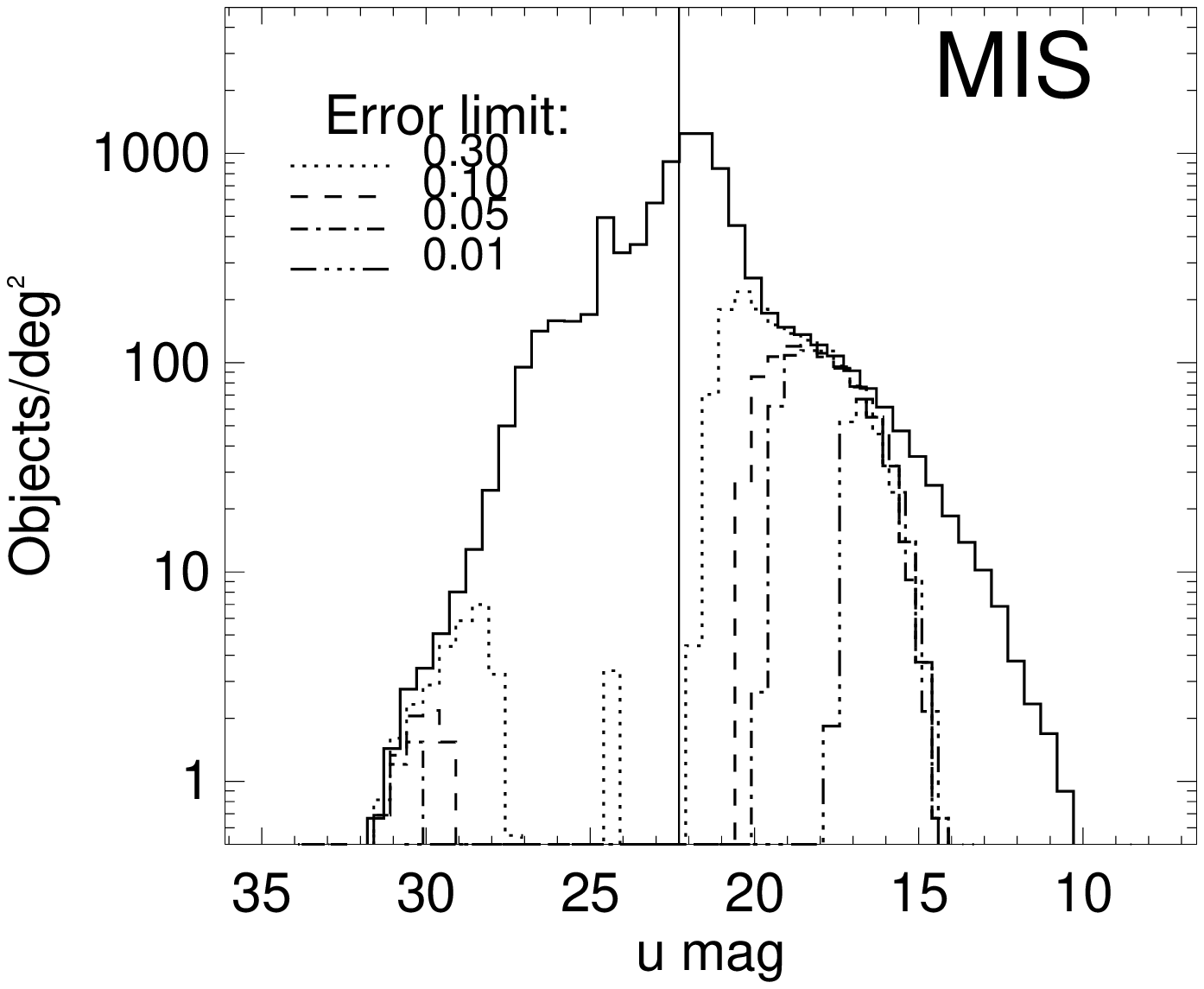,width=0.5\hsize,clip=}
\psfig{file=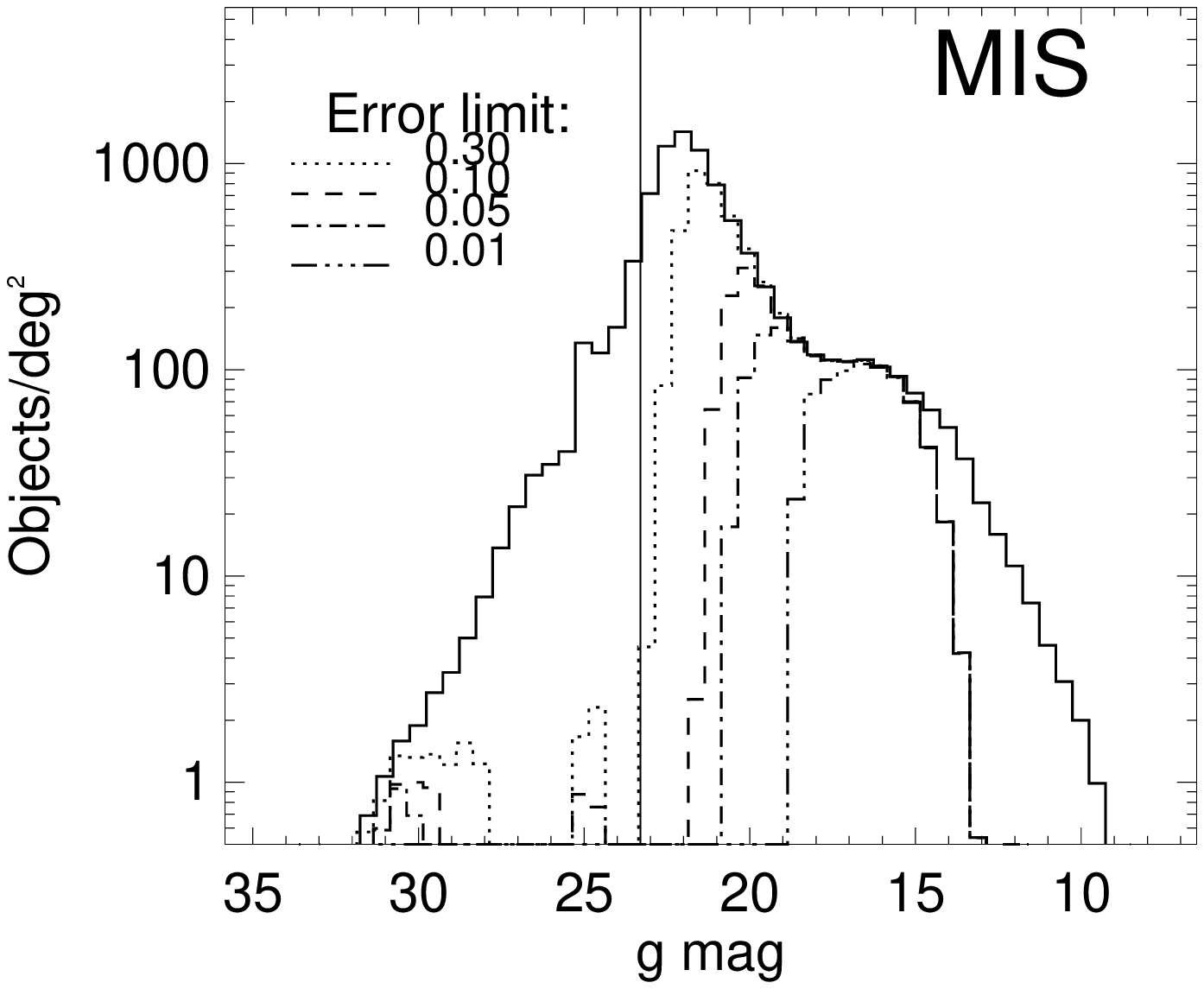,width=0.5\hsize,clip=} 
}\centerline{
\psfig{file=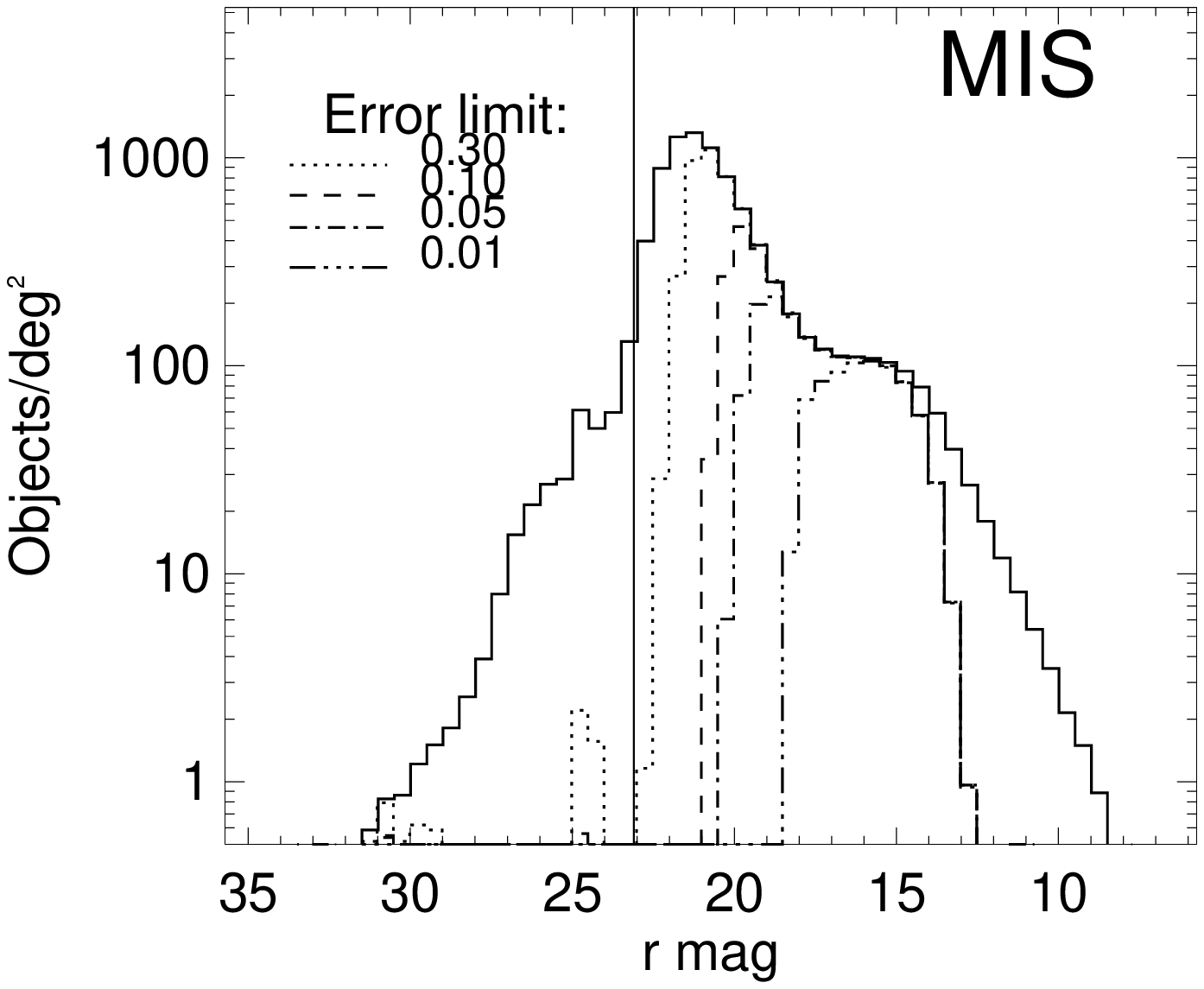,width=0.5\hsize,clip=} 
\psfig{file=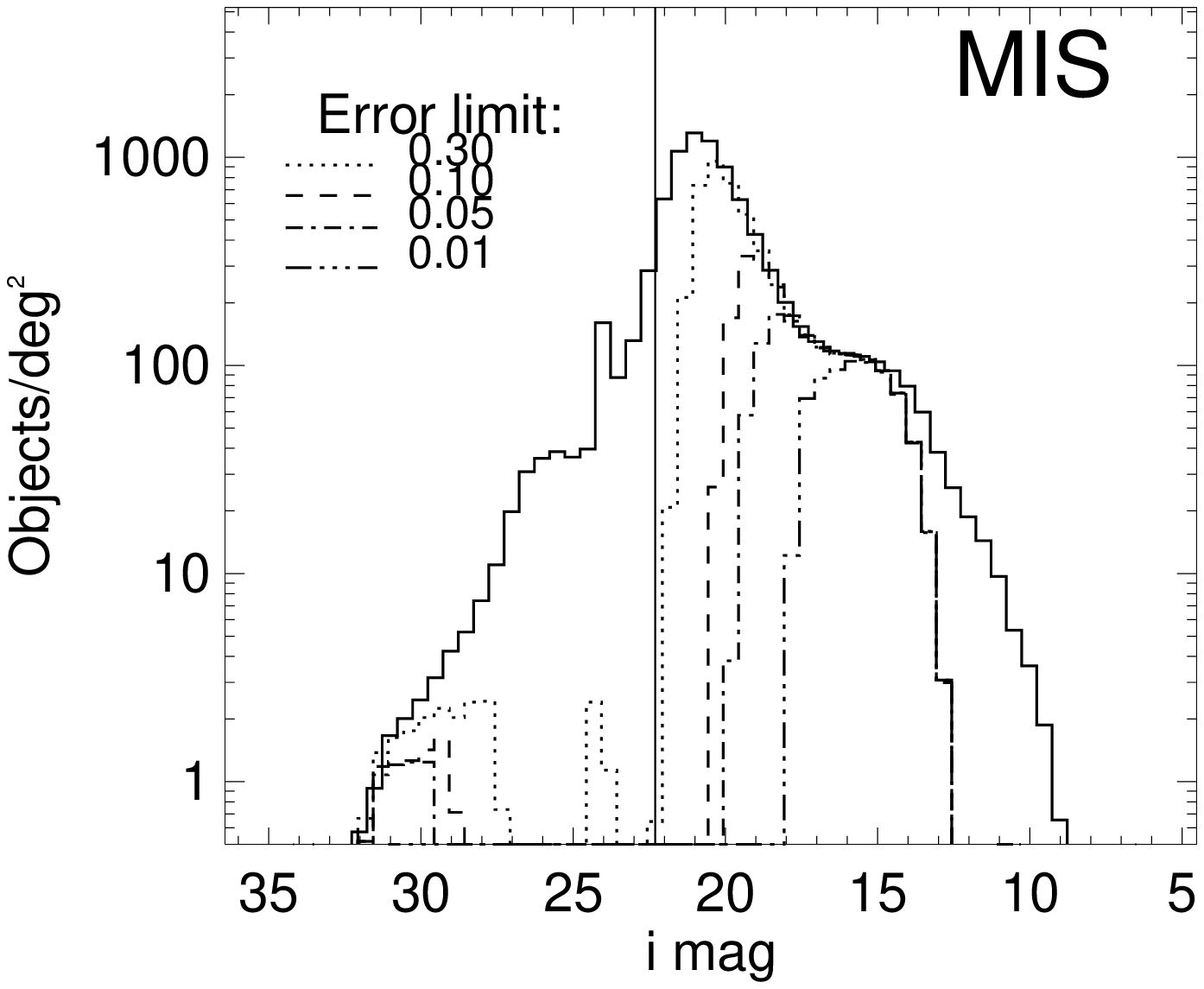,width=0.5\hsize,clip=} 
}
\caption{Similar to Figure \ref{plot_hist1} for the SDSS filters.
Only the sources matched to the GALEX MIS sources are shown, since the depth 
of the SDSS DR3 is obviously the same for the AIS matched catalog, but the
catalogs are limited by the GALEX detection limits in the AIS. 
Vertical lines indicate the limiting magnitudes in each band from  the SDSS
web site. 
The solid line is the total number of objects 
(including objects with saturation). The other lines show the histograms
when error cuts are applied.}
\label{plot_hist2}
\end{figure}

\begin{figure}[!ht]
\centerline{
\psfig{file=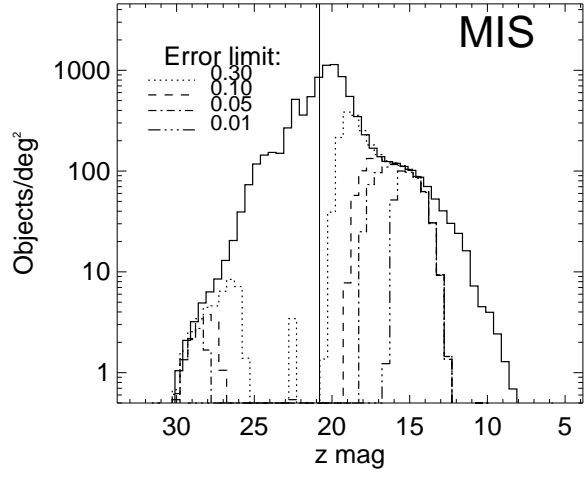,width=0.5\hsize,clip=} 
}
\caption{Similar to Figure \ref{plot_hist2} for SDSS {\it z} magnitudes.}
\label{plot_hist3}
\end{figure}

Figures \ref{plot_hist1}--\ref{plot_hist3} display the number of objects 
in our matched catalog
as a function of  magnitude for each band. The histograms  show the 
total number of objects included in the final catalog (all detections), 
as well the numbers obtained by applying different error cuts.  
These histograms are useful for estimating 
 the magnitude limits reached in restricted
samples (by applying error cuts in the analysis),
which are helpful for interpreting the results. 
In the histograms, we also marked the typical limiting 
magnitude for each band 
as given by the projects via the MAST and SDSS 
web sites. Note that the SDSS pipeline  generates  sources several 
magnitudes below the nominal limits for each band, 
and it is surprising that the error cuts are not sufficient 
to eliminate these spurious detections. 
Figures \ref{plot_hist2} and \ref{plot_hist3} show that while the 
histograms of sources brighter than the 
 magnitude limit behave as 
expected, with fainter sources progressively eliminated by 
more   stringent error cuts, there is an 
almost specular distribution of spurious sources at the faint magnitudes, 
well below the expected SDSS limit. 
Such spurious sources are different in each band, therefore while
they are not eliminated by error cuts in an individual band,
they are  eliminated   by applying both error and magnitude cuts to the sample.

In the following section, we 
discuss the effects of 
  error cuts applied as 
appropriate  to separate the sources into classes of astrophysical 
objects based on  their colors, in order to 
minimize contamination by other types of objects. 
Error cuts translate (statistically)  into magnitude limits, 
in each photometric band, as can be inferred 
from the histograms in Figures \ref{plot_hist1}--\ref{plot_hist3}. 
The number of objects remaining in our catalog when different 
error limits are applied in each band
are given in Table \ref{tbl-1}. 
It is extremely important 
 to  consider the effects of the magnitude 
limits when using colors to classify objects and to derive 
luminosity functions. Therefore,  we provide some  illustration of such
 effects and discuss them 
in section 4.

In our analysis we also   use SDSS spatial information 
to separate the sources into point-like 
(at the SDSS $\sim$1.4$''$ psf) and extended sources. 
Point-like sources are mostly stellar objects 
or QSOs (see section 4), while extended sources are typically galaxies.  
 The numbers of point-like and extended 
sources are approximately equal at the depth of the AIS survey, 
but the number of extended 
sources (galaxies) increases significantly at fainter magnitudes 
(MIS survey) as shown
in Figures \ref{plot_cc1} - \ref{plot_cc3}.
 Likewise, among the point-like sources, the low-redshift QSO candidates
increase more  than the number of hot star candidates 
at fainter
magnitudes (section \ref{s_color}). 

\section{Analysis. Classification of the GALEX sources.}
\label{s_analysis}

In this section we analyse the colors of the matched sources to extract 
information about their nature.  
Bianchi et al. (2005) showed two sample color-color diagrams 
demonstrating that the combination of the  GALEX UV bands
with optical photometry  is a powerful tool for identifying
 low-redshift QSOs, 
hot stellar objects and binary stellar objects containing a hot WD.

\subsection{The model colors}
\label{s_models}
\begin{figure}[!ht]
 \vskip -2.cm
\centerline{
\vbox{\hbox{
\psfig{file=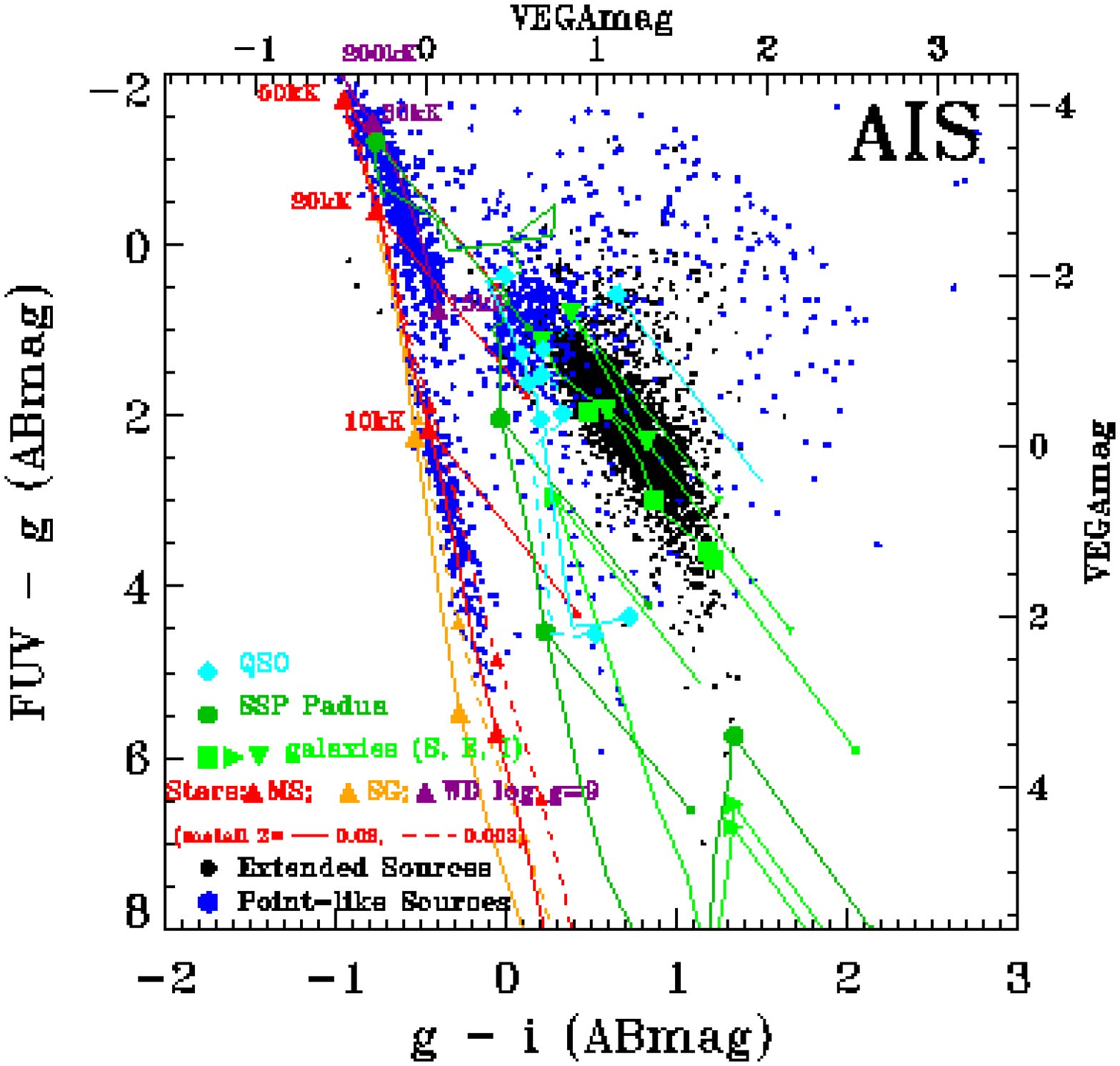,width=3.5in}
\psfig{file=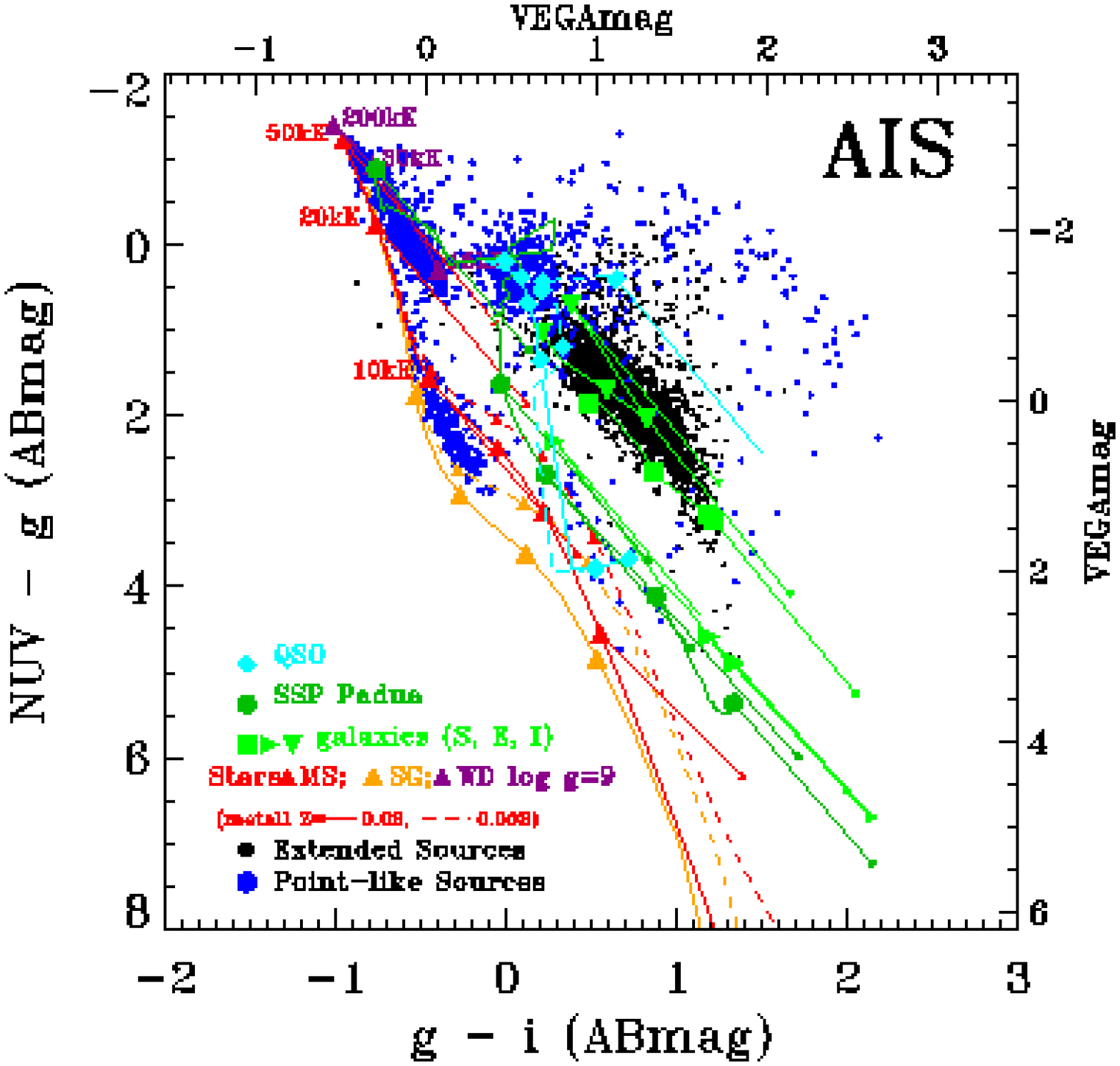,width=3.5in}
}
\hbox{
\psfig{file=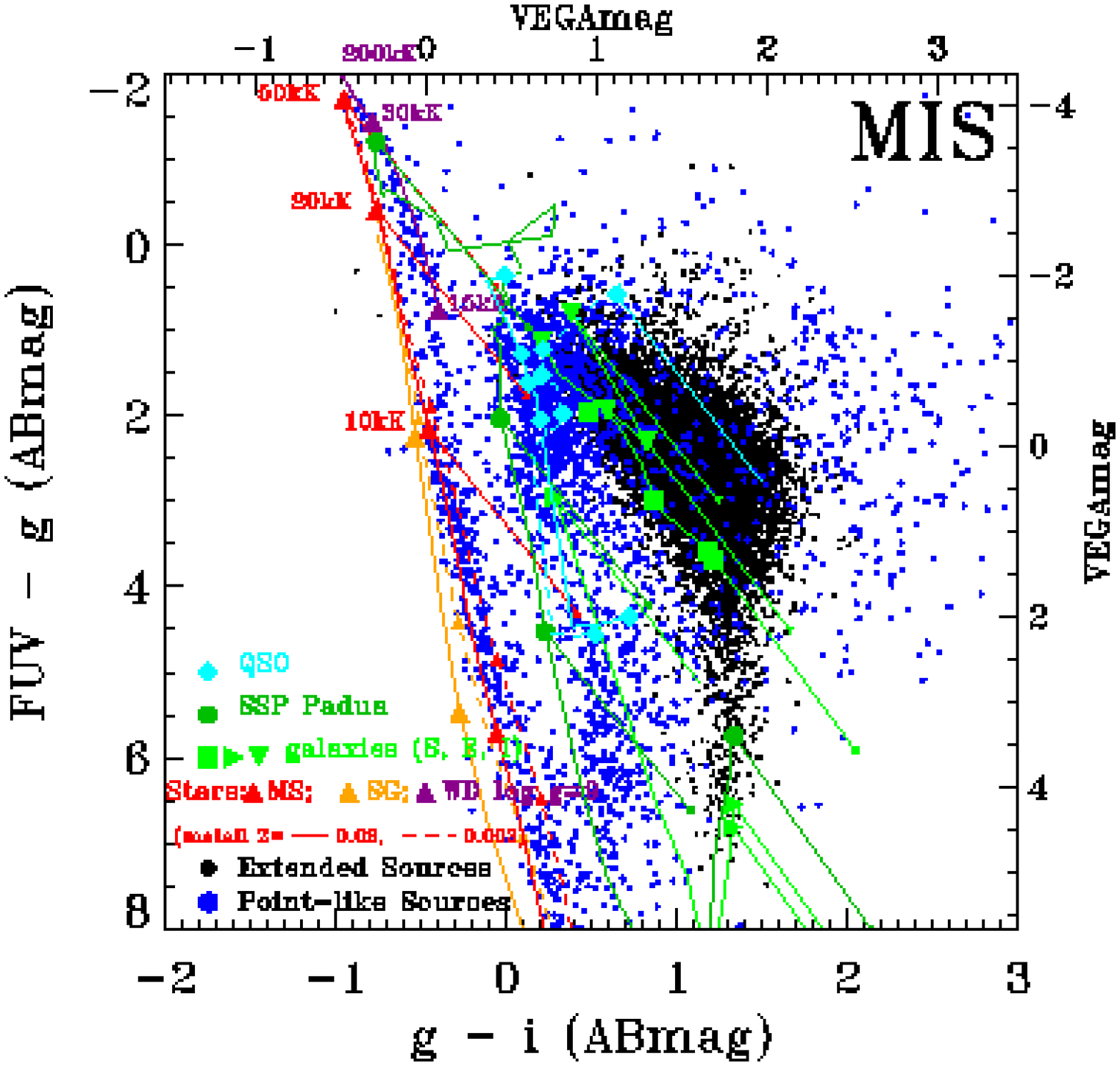,width=3.5in}
\psfig{file=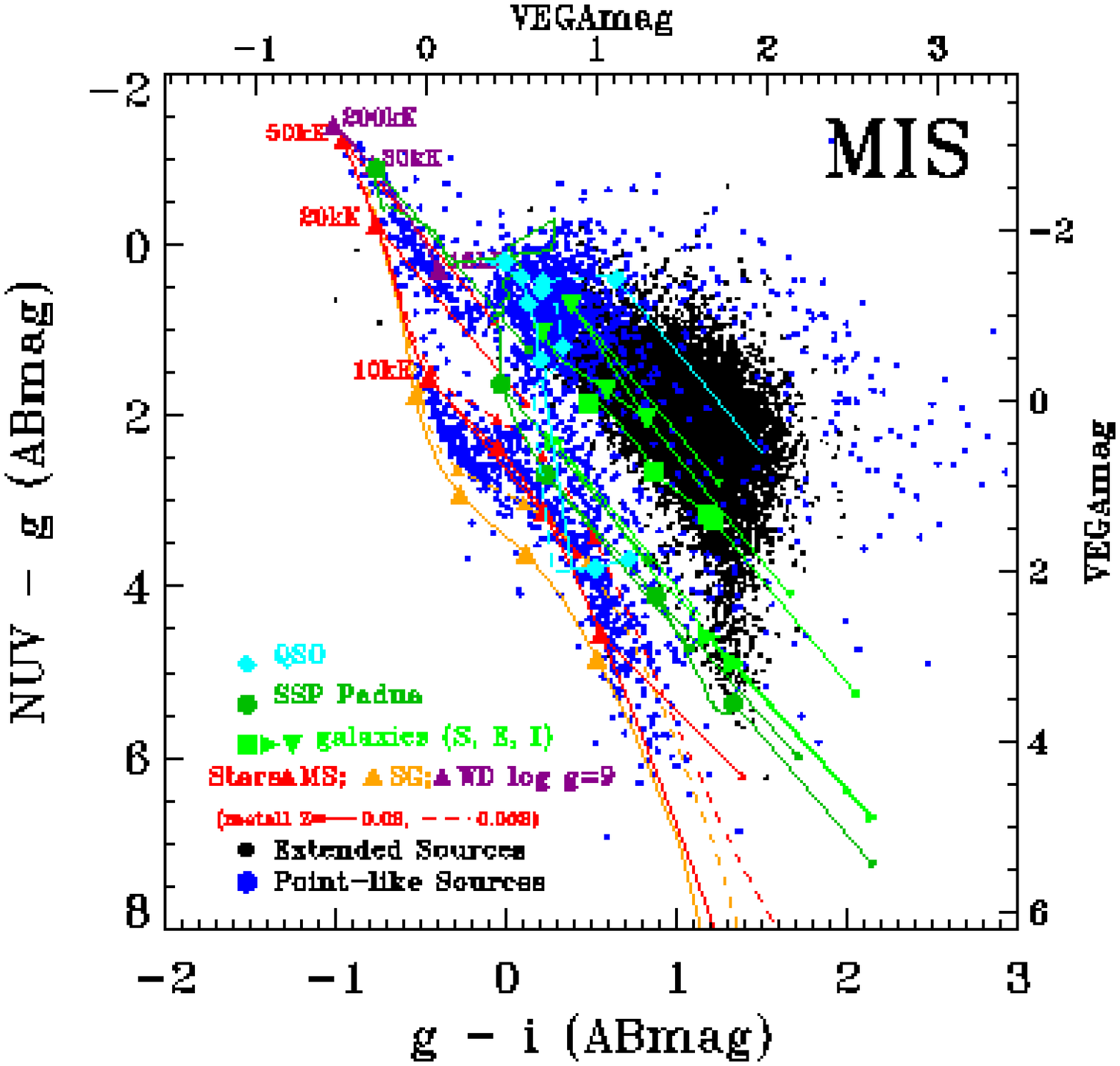,width=3.5in}
}
} }
\caption{ \footnotesize The \fggi and \nggi color-color diagrams for the AIS and MIS sources. 
Blue dots are point-like sources
and black-dots are extended sources, based on the optical data (psf $\approx$ 1.4''). 
Model symbols are explained in the legend and described in the text.
[FUV-g] and [NUV-g] are shown on the same scale, to illustrate (y-axis)
the advantage of FUV for separating hot stellar objects. 
Extinction effects are shown for \ebv=0.5\,mag, MW-type dust, with
lines connecting large symbols (intrinsic model colors) to smaller symbols 
(reddened model color). 
Triangles mark stellar (main sequence and supergiants) \Teff's of 
50, 20, 10, 8, 7 and 6 kK.
 QSO model symbols mark redshifts of {\it z}= 0.0, 0.6, 1.0, 1.6  and 3.
Galaxy templates symbols (green) mark ages of 1, 5, 12 and 13 Gyrs for
Ellipticals (E), Spirals (S), Irregulars (I). 
SSP marked ages are 1, 200, 500Myrs, 1.5 and 15Gyr.
Point-like sources cluster along the stellar sequence 
(predominantly on the  high-gravity sequence at high \Teff's) and the 
 QSO locus.
QSOs and galaxies
occupy contiguous loci, well separated from the stellar sources.   
Sources with photometric errors better than 
0.15/0.10/0.05/0.05\,mag in FUV/NUV/g/i are shown. 
}
\label{plot_cc1}
\end{figure}

In Figures \ref{plot_cc1} -- \ref{plot_cc3} we  compare 
observed colors of our sources 
 to model colors, 
calculated by applying the transmission curves of the GALEX and SDSS filters
to theoretical spectra or templates  of different 
astrophysical objects.
Theoretical stellar colors for main sequence and supergiant stars
are derived from the stellar 
libraries of 
Lejeune (1997), Bianchi et al.  (in preparation),
and Rodriguez et al. (2005), for a total of six metallicity values
(two cases shown in the color-color diagrams: solar, and one tenth
solar). All these grids are based on Kurucz 
(LTE, line-blanketed, plane-parallel) models. 
We compared, for sample cases,  the  broad-band stellar 
colors derived from the Kurucz model grids, to 
colors computed from TLUSTY (Hubeny \& Lanz 1995) 
 model  atmospheres for hot stars 
(non-LTE, line-blanketed, plane-parallel), as well as  with
stellar atmospheres plus wind models
(non-LTE, line-blanketed, spherical, hydrodynamic), 
computed by us  (Bianchi \& Garcia 2002, Garcia \& Bianchi 2004, 
Bianchi et al., in preparation) 
with the WM-basic code of Pauldrach et al. (2001).
As expected, there is no appreciable difference 
in the broad-band colors. The whole grid of models will be published
elsewhere (Bianchi et al., in preparation). 
We calculated white dwarf (WD) models 
 using the TLUSTY code of Hubeny \& Lanz (2005), 
 for various gravities
and solar, pure-H and pure-He abundances. 
In the color-color diagrams (figures \ref{plot_cc1} -- \ref{plot_cc3}),
 we show stellar model colors as a function
of \Teff ~ for three different gravities (representing main-sequence (MS),
supergiants (SG) and high-gravity stars with log~g=9), and   
two metallicity values, to illustrate the effects of relevant 
parameters  without crowding the figure.

To define the locus of galaxies, we computed broad-band colors from
``Simple Stellar Population'' (SSP) models 
from the library by Bressan et al. 2006 ({\it in preparation}), 
as well as from a library of galaxy templates.
The templates were calculated for ellipticals, spirals and irregulars 
(marked as ``E, S, I'' in the figures)
using the GRAZIL code (Bressan et al. 1998, Silva et al. 1998). 
We show in figures \ref{plot_cc1} -- \ref{plot_cc3}
these model colors  
as a function of age, for redishift {\it z}=0. 
 For QSO colors, we used a composite model of Francis et al. (1991)
 revised as described in Bianchi et al. (2005), 
as well as the SDSS average template, with extrapolations at long and short
wavelengths by Zheng (private communication),  that include Ly$_{\alpha}$ absorption.
Model colors from both these QSO templates are shown as a function of
red-shift in the color-color diagrams, the first plotted with a solid
line, the latter with a dashed line.

Bianchi et al. (2005) showed two sample  
color combinations, for the GALEX early data release sample, 
and also plotted a comparison of [known] spectroscopically 
classified objects with  the model colors, showing that the {\it loci} 
defined by our models match very well the observed 
properties of each astrophysical class represented. 
We omit the comparison of the models to known objects in this 
work as it would show the same result. Instead, in this paper
we describe 
the effects of physical parameters, such as 
\Teff, metallicity and gravity
for stars.   Below we discuss general properties and then 
we comment on individual 
color-color diagrams.

Effects of interstellar extinction by dust are applied
to the model spectra, for varying amounts and types of extinction,
in the calculation of the grids of model colors.  
Because we are using broad-band photometry, applying extinction
corrections A$_{\lambda}$/\ebv ~ for the ${\lambda}_{eff}$ of 
the filter band-pass to the intrinsic model magnitudes 
would only be an approximation of the reddening effects on the colors.
Instead, we reddened each model spectrum with progressive amounts
of \ebv , for different extinction laws, and then calculated
broad-band colors for the reddened model spectra.   For the analysis
of hot Galactic objects, and for the sample in this paper in particular, 
which includes fields at  high galactic latitudes, 
the extinction is small as we shall see in the next section. 
Therefore, we only
show the case of Milky Way-type extinction with R$_V$=3.1 in the 
color-color diagrams. 
    
\subsection{Color-Color diagrams}
\label{s_color}
Figures \ref{plot_cc1}--\ref{plot_cc3} show a number of 
color-color diagrams of the sources and model colors, chosen  to illustrate
the most relevant effects of the physical parameters  and useful
classification criteria, as well as the statistical
properties of the sources.
In these figures, 
point-like sources are shown 
with  blue dots  and extended sources with black dots. 
To avoid excessive crowding in the diagrams, in general
only sources with
errors better than 0.15\,mag (0.10\,mag) in FUV (NUV) and
 better than 0.05\,mag
in the optical bands are shown. More details
are given in the captions.  
Model   colors are shown with different 
symbols and colors, explained in the figures. 
 The effects of 
reddening by interstellar dust are shown with lines
connecting the intrinsic model color (large symbol) to the reddened
model color (smaller symbol), 
for a sample case of Galactic typical 
extinction with  $R_V$= 3.1, and for  E(B-V)=0.5\,mag. 
It can be seen from all diagrams that for most  hot stars
the probable reddening does not exceed \ebv $\approx$ 0.1\,mag, as can be
expected given the high galactic latitudes of the fields
included in our catalog (Figure 1). 

\subsubsection{General Properties}

\begin{figure}[!ht]
\centerline{
\vbox{\hbox{
\psfig{file=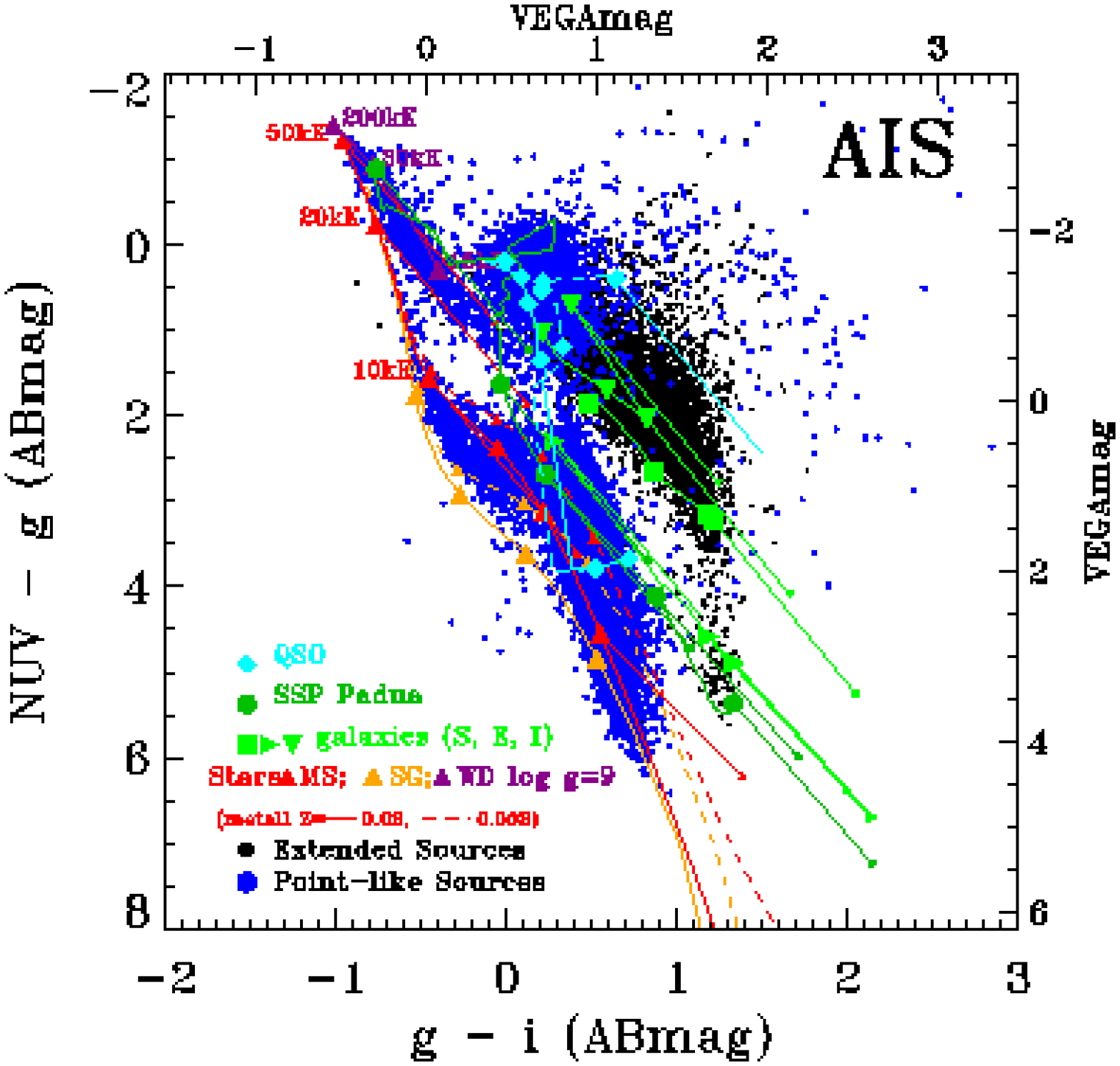,width=3.1in}
\psfig{file=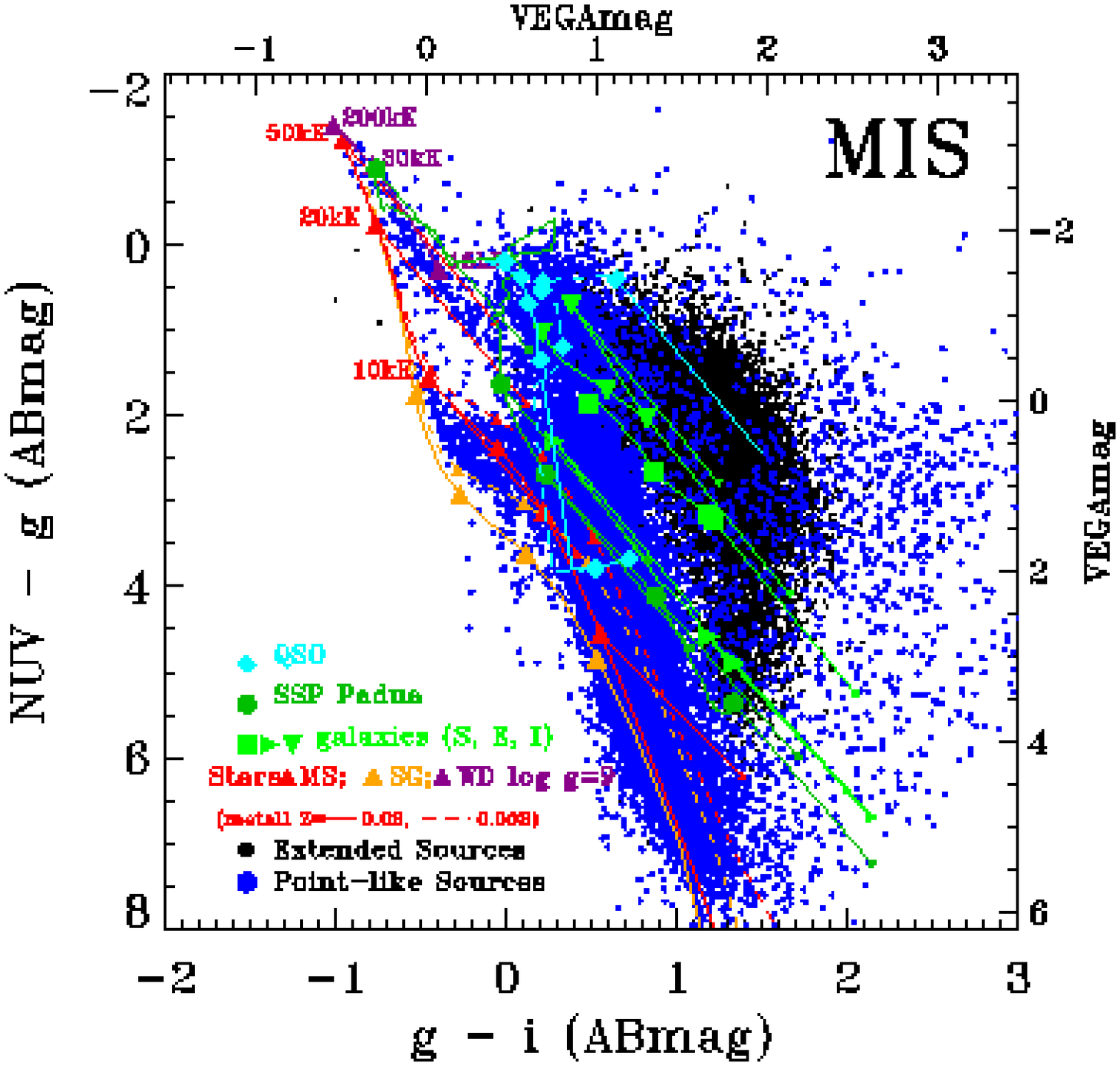,width=3.1in}
 } } }
\caption{The  \nggi color-color diagrams for the AIS and MIS sources,
with symbols as in the previous figure. 
In this figure 
 sources with errors $<$ 0.1\,mag in NUV and 
$<$0.05\,mag in the optical bands are shown, regardless of 
FUV-band photometric error. 
The comparison with the previous figure, where an error cut 
in the FUV band was also applied, illustrates the selection  effects
driven by FUV magnitude limits and error cuts in the analysis.
While the hot objects hardly change, the lower \Teff~  stellar sequence 
 becomes extremely populated in these panels. 
}
\label{plot_cc2}
\end{figure}

In all the color-color diagrams, but especially in the
\fngr  diagram (Figure \ref{plot_cc3})
and \fggi (or \nggi ) diagram (Figures \ref{plot_cc1} and \ref{plot_cc2}), the extended sources (black
dots, mostly galaxies) are well separated from the stellar sources
(the blue dots, or point-like sources, near the stellar model colors). 
The {\it locus} of the QSOs lies somewhat in between, and is occupied in these
diagrams by  a compact cloud of mostly point-like sources,
populating the {\it locus} of the low-{\it z} QSO template colors.
A number of extended sources is also found in the same color space,
which is also shared by the young-age SSP and  elliptical galaxy models. 
Note that if we consider points with good photometry also in
the FUV band, only  the locus of low-{\it z} QSOs is populated
with sources. Instead, sources with good photometry  in the 
NUV---optical 
range (FUV drop-outs) do extend over the color-space of higher
redshifts QSOs, as expected, and of stars with lower temperatures.   

Another general feature which is  evident in all diagrams, from
the density of points in regions defined by  the model colors for 
different   objects, is that the
number of extragalactic sources (galaxies and QSOs) increases
much more than the number of Milky Way stellar sources at
fainter magnitudes.  This effect is quantified by the 
surface density plots shown in the section 5. 

The galaxies (extended sources, black dots) are well represented
by the three ``average'' templates for irregulars, 
spirals and ellipticals. Two properties of the sample can be
inferred  from the color-color diagrams.  First, the 
templates cover well the space for irregular and spiral
galaxies in the AIS sample, which is limited to brighter magnitudes.
For the deeper MIS  sample the 
templates' intrinsic colors rather provide an envelope to 
distribution of the sources,
and the extinction effects shown (foreground MW-type) do not account for
the entire extent of the data-points on the color-color diagrams. 
This may be an indication
of intrinsic mixed, possibly non-MW-type extinction.
The analysis of galaxies colors from the GALEX surveys 
is not the scope of this work and is pursued in other works,
therefore for our purpose it is sufficient that 
the unreddened galaxy colors produce an envelope to
the observed distribution of the extended sources. 
The second point of relevance
is that data-points are seen with colors corresponding to 
Ellipticals of either young or old ages, 
due to the contribution of hot He burning low mass stellar
population such as
       horizontal-branch stars and their progeny
  (Yi, 2003) 
in the diagrams including the FUV band, in which  these objects
are probably too faint at intermediate ages. This is especially evident 
in the shallower AIS sample. When the more sensitive NUV band 
is considered (regardless of FUV detection of the source),
the locus of Ellipticals, especially at old ages, becomes
more populated, even in the AIS survey (Figure \ref{plot_cc2}). 

As for the stellar sources, our models for different 
luminosity classes 
(main sequence versus supergiants) and metallicities 
show the separation of the photometric colors
for stars cooler than approximately 10,000~K, most
evident in the \nggi ~ diagram. For hotter stars,
supergiants and main-sequence colors are indistinguishable,
however the well defined cluster of stellar  sources separate
from the main-sequence colors is well represented
by high gravity stellar  models. Only WD models with log~g =9 are shown
for clarity. 
We see that the majority of the hottest stellar sources lies
on the high-gravity sequence (evolved stars) rather
than on the main-sequence or supergiant color sequence. 
An interesting point is that, for the case of MW-type extinction
(as shown),  reddening would displace the intrinsic color
of a very hot star roughly along 
 color sequence of high-gravity stars
in the \nggi ~ diagram,
making \Teff~ and reddening effects indistinguishable. 
However, the effects of \Teff~ and reddening on the stellar colors are different
in the \fngr and \fggi diagrams and they can be disentangled. 
In any case, it is evident from all diagrams that the number of
objects along the high-gravity stellar sequence increases at lower
\Teff 's, consistent with numerical expectations. It would be 
unlikely that the majority of objects were extremely hot, 
highly reddened main-sequence or supergiant stars. 

Another  interesting stellar population component that can be easily detected,
especially  in  the   \fnnr ~ diagram  
(Figure \ref{plot_cc3}), is that of stellar 
binaries composed of  a white dwarf and a much cooler object.
 A significant number of  point-like sources can be
seen with very blue FUV-NUV colors, but red optical colors.
These objects appear as an arc-like distribution of blue 
points on the top-right part of the color-color diagrams, 
and are presumably  WD plus M-dwarf pairs. 
The WD+MD binary sequence has been first recognized at optical
colors (e.g. Raymond et al. 2003, Smolcic et al. 2004, 
Pourbaix et al. 2004) but the UV bands are much more sensitive
to the detection of systems with hot WD.  
Smolcic et al. (2004) found   
$\approx$ 0.4 of such binaries per square degree from
the SDSS DR1.  
This population will be discussed again in section \ref{s_disc_hot}. 

The last general point worth discussing, and often overlooked,
is the effect of the photometry's magnitude limits (and hence
also effects of error cuts) on the sample. 
Figures \ref{plot_cc1} and \ref{plot_cc2} show the same
\nggi ~ diagram, in one case selecting sources with good
photometry in the four filters shown, regardless of detection
in the FUV (Figure \ref{plot_cc2}). In 
Figure \ref{plot_cc1} the same sample is restricted to sources
that also have good photometry (error better than 0.15 mag) in the
FUV band. As expected, the hot star, and low-redshift QSO
samples do not change significantly, however QSOs at higher redshift and a
huge number of cooler stellar sources appear 
only when no FUV cut is applied. 
Interestingly, also a number of extended sources occupy the crowded locus
of the lower \Teff ~ stellar sequence and higher-{\it z} QSOs. 
These data-points are not visible in the diagram because the
point-like sources (the primary subject of investigation in this work)
are plotted over the extended sources. 
In general, imposing good photometry also in the FUV band reduces
the sample by a factor of about 10, with respect to a sample 
based on NUV-band good photometry (see also Table 1).
 Obviously, 
magnitude limits  affect objects differently depending
on their colors.

\subsubsection{Discussion of the individual color-color diagrams}

We finally comment here on each diagram separately.

The \nggi and \fggi
color-color diagrams (Figures \ref{plot_cc1} and \ref{plot_cc2})
are somewhat similar to the classical optical {\it [U-B] vs [B-V]}
diagram (e.g. Bianchi et al. 2006a) but present several advantages 
thanks to the broader wavelength range. Stars are well separated 
from  galaxies and low-redshift QSOs in the {\it g-i} color. 
Binary stars containing a hot WD and a much  cooler companion
are easy to detect (top right blue points in the diagram). 
The price to pay for the wide color separation of the objects
is that the very hot WDs are faint in the {\it i} band, thus
a sample selected from these colors is clean but limited to
 objects brighter than the detection depth that can be reached by the GALEX surveys
(see next section). We plot the {\it FUV-g} and {\it NUV-g}
colors (left, and right panels respectively) on the same scale,
to illustrate the advantage of including a FUV band as diagnostic
for the hottest objects: note the much larger spread in color
among hot temperature models, when the FUV band is included.
 Broad-band photometric colors separate
for stars of different gravity and metallicity only at \Teff~ 
lower than $\approx$ 10,000K, as shown  also
by Bianchi \& Efremova (2006) in a recent HST study of stellar populations. 
For high-gravity stars, we plot only one
sequence of model colors with log~g=9 for clarity.
While supergiants and main sequence stars 
 are indistinguishable at high \Teff s, gravities
much higher than log~g=5 (our model sequence plotted in red) are clearly 
separated photometrically.
Most of the GALEX hot stellar sources cluster around the log~g=9 
sequence and  are therefore likely to be  subdwarfs or WDs. 
A dramatic difference in the sources' census is made by including or
neglecting the FUV photometry (Figure \ref{plot_cc1} versus
\ref{plot_cc2}); this effect has already been mentioned above. 
Finally, note the direction of the reddening effect on the colors in the 
\nggi diagram: it is very similar to the effect of the major
physical parameters (\Teff~ for stars, age for galaxies, 
redshift for QSOs - in some ranges), making these particular colors unsuitable
to disentangling reddening and e.g. \Teff~ for WDs.  
The degeneracy is less severe for the \fggi  diagram, and it
is completely removed in the diagrams including a [FUV-NUV] color
(Figure \ref{plot_cc3}). Such extinction effects apply only to the case
of MW-type dust, and would be slightly different for other types of
dust.

Similar  considerations apply to 
 the \fngr diagrams (Figure \ref{plot_cc3} , top),
however stars with different gravities are less separated, 
while galaxies, QSO and stars are better separated, with
respect to the diagrams discussed previously.  

The NUV-{\it r} vs. FUV-NUV diagram (Figure \ref{plot_cc3}, lower panels)
 separates very well stellar objects from galaxies, thanks to the broad 
wavelength base-line. The stellar 
sequence is continuously populated, with the number of stellar objects 
fading out at \Teff~ about 
 8500~K  in the AIS catalog, but extending to cooler (\Teff $\sim$ 6000~K ) stars in the deeper 
MIS catalog. Note that these limits are driven by imposing an error
cut of FUV$_{err}$ $<$ 0.15mag (as used in the color-color diagrams). 
In particular, hot stellar objects can be easily selected from their FUV-NUV color,
while  a UV-r color allows us to separate single hot stars from binaries 
with a hot and a cool  component.  
Using the NUV-{\it r} and  FUV-NUV colors we extract hot star candidates in different \Teff ~ ranges,
and low redshift (z $<$ 1.6) QSO candidates (next section). 
 The distribution of hot stars and QSO candidates
are shown  in Figures \ref{plot_LF_WD} to 
\ref{plot_LF_QSO}.

In the next section we present the statistical properties of 
the hot stellar objects
(single and binary) candidates. 
 An analysis of the physical parameters of these objects
will be the subject of future papers 
 with spectroscopic follow-up.

\section{Discussion and Conclusions}
\label{s_disc}
As discussed in the previous section, the  color-color diagrams can be used 
to separate objects by astrophysical  class, and in particular to
select hot stars and low-redshift QSO candidates.  
In this section we derive  surface densities
 for such objects, and compare them to
previous known catalogs, as well as to
predictions by a Milky Way model. 

\subsection{Hot stars Candidates}
\label{s_disc_hot}


According to the fluxes of  our TLUSTY models, a WD with
 \Teff = 50,000K[100,000K], log g =7.,
 and R=0.2 \Rsun~ has an AB  magnitude
(in absence of reddening) of 
{\it r}= 8.5, 13.5, 18.5 and 20.0 [7.9, 12.6, 17.9, 19.4]\,mag 
at 100pc, 1kpc, 10kpc and 20kpc respectively.
Again from our models, its colors are FUV-{\it r} = -2.35[-2.62]
  and NUV-{\it r} =-1.75[-1.93] 
(ABmag) for   \Teff = 50,000K and 100,000K respectively.
A radius of R=0.2 \Rsun~ would be a typical value for a very hot 
 post-AGB star (\Teff$\approx$ 100000~K) at the end of
the constant-luminosity phase. A lower mass/ cooler remnant 
may have a larger radius. 
A hot WD then would dim by several magnitudes  descending
(at approximately the same \Teff, initially)  along the cooling sequence,
down to a radius of $\approx$ 0.05 - 0.02\Rsun , becoming fainter 
by -2.0  in
log Flux, or 5 magnitudes. 
Therefore, with  error limits 
of $<$0.15/0.3mag (FUV) and $<$0.10/0.3mag (NUV) which translate into
GALEX magnitude limits of $\approx$19.5/21.5~mag  
for AIS and 22.5/25~mag  
for MIS, we expect that our GALEX dataset can detect WDs all the way
 in the Galactic halo, 
even in the AIS sample, in their (advanced) post-AGB constant luminosity phase.
In the MIS sample, we expect to also detect hot WD with radii as small as 0.04\Rsun~
out to 20kpc. When descending on the WD cooling sequence,
the stars will have higher gravity. For log~g=9., 
    FUV-{\it r} = -2.33[-2.61]
  and NUV-{\it r} =-1.73[-1.92] 
(ABmag) for 
 \Teff = 50,000K and 100,000K respectively, not significantly
different from the log~g=7 case. 
 Imposing an error limit of better than 0.1mag 
in the {\it r~} and {\it g~} bands would however  limit the survey to a smaller volume (see 
Figure 3). 
These model magnitude values  are in absence of reddening. However, as we saw
from the color-color diagrams, reddening is quite small for the
high galactic latitudes of the present sample. 

For subdwarfs and main-sequence stars, e.g. with log~g =5 and
\Teff = 50kK, 30kK, and 18kK, 
FUV-{\it r} = [ -2.30, -1.87 and -0.55], 
NUV-{\it r}= [-1.74 ,-1.38, and -0.40]. The radius will be
larger than that of a WD, therefore all hot stars are expected to be detected in
our GALEX surveys.  For a reference radius of 1\Rsun ,
at 20kpc a star with log~g=5. and   \Teff = 50kK,  30kK, 18kK would have
{\it r} = 16.53, 17.34, and 18.26mag.  These numbers are based on 
our TLUSTY model calculations.  Using Kurucz models for log~g=5 stars,
we find consistent magnitudes within 0.1\,mag.

Here we use two different color  selections to extract hot stars candidates
from the GALEX sources.  First, 
the color-color diagrams \nggi or \fggi (Figure \ref{plot_cc1}), 
and  the similar \nggr and \fggr, 
can be used to select hot stars. At the hottest temperatures, 
main sequence and supergiant stars cannot be separated, but high-gravity (WD) stars
are well separated and are the majority of the hottest sources.
At lower \Teff , MS stars and TO stars dominate the statistics (Figure \ref{plot_cc1}). 
 Figure \ref{plot_LF_WD} shows the number per square degree  of high gravity  
stars, and lower gravity stars, 
selected from the \nggr colors, with error cuts of 
0.1mag (NUV, {\it r} and {\it g}). 
If we  included sources with larger errors, they would also be consistent with  the
locus of QSOs in the  \nggr and \fggr diagrams. 
 In this selection,
hot stellar sources with a cool companion (occupying the top-right
part of the color-color diagrams) cannot be included, because
for {\it g-r}$>$-0.1 they would be confused with QSO candidates.
Therefore, our selection included the locus around the stellar sequence
for high gravity (purple symbols in figures \ref{plot_cc1}) 
down to \Teff$\approx$15000~K for WD. For the 
lower gravity stars, 
we included
all the lower \Teff~  sequence and no {\it g-r} cut below \Teff=10,000K,
so binaries are also counted in this case. 

The error bars on the number densities  in Figure \ref{plot_LF_WD}
are calculated by considering combined
photometric 1$\sigma$ errors of each object, and counting
the number of sources that fall inside and outside the
color boundary that define an object class, when their individual
errors are applied in both directions.


 In  Figure \ref{plot_LF_WD} we also plot predictions for the
surface density
 of such objects calculated with the Besan{\c c}on Milky Way model  
 (Robin et al. 2003, 2004).  
We assumed a standard diffuse absorption
of A$_V$=0.7/kpc, and calculated the predictions over an area 
of approximately 210 square degrees between 40~$\deg$ 
and 50~$\deg$ in galactic latitude.  No population age
selection has been applied. The calculations were extended
to {\it g}= 24. 
The Milky Way model predictions are
 shown as black histograms, with symbols
of the same color of the objects they represent (green for WD,
blue for MS and TO, and orange for total) in the data histograms. 
The lower than predicted number of bright objects is due to
saturation in the SDSS bands (saturated objects are excluded in our sample)
which occur around 14th mag (between 13 and 15 mag).  At the fainter
magnitude end, incompleteness is due to the error cuts.  



We also performed a selection of hot WD candidates based on the 
[FUV-NUV] color only, down to \Teff $\approx$ 18,000K.  
Below this limit their FUV-NUV color   
 overlaps with the {\it z}$\approx$0 QSOs; 
our QSO templates
with {\it z}=0 have FUV-NUV=0.175~mag. 
We  used [FUV-NUV] $<$ -0.037~mag (corresponding 
to \Teff $>$ 18,000K for log~g =5)
and  [FUV-NUV] $<$-0.343~mag, which 
corresponds to \Teff $>$ 30,000K for log~g =5
but includes objects down to \Teff $\approx$ 25,000K for higher gravities. 
The surface density  of the hot stars selected in this way, 
from GALEX measurements only (errors $<$ 0.3~mag), 
is shown in  Figures \ref{plot_LF_HS} and \ref{plot_LFNUV_HS}.  
If we  restrict the sample to sources with 
 err$_{FUV}$$<$0.15~mag  and err$_{FUV}$$<$0.10~mag, 
the shape of the histogram
 does not change significantly, but the density of objects 
becomes about half in both AIS and MIS, and the counts begin
to drop at about one magnitude brighter than they do in 
Figure \ref{plot_LF_HS}.

The WD surface density  in Figure \ref{plot_LF_HS} 
is matching the  predictions by the Milky Way model
in the MIS sample, and is a factor of two  lower
in the AIS sample. 
However, we suspect that the MIS counts are contaminated by point-like
sources possibly of extragalactic nature, as explained and quantified below.
Such spurious sources  affect only the ``binary'' candidates, and
not the ``single WD'' candidates, as we will see.

 The number of objects per square degree (surface density)
are plotted 
 in Figures  \ref{plot_LF_WD} and  \ref{plot_LF_HS} 
 as a function of the
visual {\it r } magnitude, for direct comparison with the Galactic model
predictions and with other works.  When we combine
the GALEX selection of hot stellar objects
(from  [FUV-NUV] color) with optical bands, we can also
assess the fraction of hot stars that have a cooler companion,
i.e. objects with FUV-NUV color corresponding to a hot \Teff, but
much redder optical colors.  However, objects with WD-like
FUV-NUV colors and brighter than expected optical magnitudes
may also be of extra-galactic origin, such as unresolved galaxies
or QSOs whose SEDs may differ from our templates.  
The fraction of 
objects with optical colors inconsistent with their
hot-WD FUV-NUV color,
 is about 20-30\% for magnitudes brigher than $\approx$20, however
the number 
substantially increases at fainter magnitudes.
This increase suggests some contamination by non-stellar
(extra-galactic) objects.  
The {\it r } magnitude of a fraction of our sample objects therefore 
does not correspond to that of the WD component, but rather 
to the optical magnitude of a cool companion, 
or to that of a different object in case of extragalactic
contaminants. In either case, the observed {\it r } magnitude is 
  brighter than the  WD optical magnitude. This fact causes
an artificial increase of the number of MIS sources 
 in the bright {\it r } magnitudes range,
where we would expect that the density of sources (either single WD, binaries
or possible spurious objects) would anyway be the same in
AIS and MIS,  down to the limit where the AIS begins to be incomplete.
Therefore, we consider the match of the Milky Way model to the MIS counts 
in Figure \ref{plot_LF_HS} (left) to be, at least in part,
 a bias from this effect. In order to avoid this bias, 
we plotted  in Figure \ref{plot_LFNUV_HS}
the surface density for the 
same sample of hot-star candidates as a function of NUV magnitude,
 which better represents the magnitude of the  WD component
in the binary objects.  The solid-color histograms show the 
``single'' objects (whose optical colors are consistent with 
the FUV-NUV hot-star classification). These histograms show
similar object counts for MIS and AIS, down to where the AIS 
becomes incomplete, as one would expect. 
The dashed histograms show all FUV-NUV selected
sources, both single- and  binary-WD candidates,
the latter including  possible spurious objects. 
These total counts increase at fainter magnitudes much more than the 
single-WD counts, reinforcing the suspicion of contamination by 
extragalactic objects in the ``binaries'' sample.
Because the Besan{\c c}on Milky Way model is not available specifically
for GALEX bands, we have taken the model prediction  in the 
{\it r}-band and translated it into NUV-band using an average
NUV-{\it r} color in the range of \Teff~ where our WD candidates were
selected,    $<$NUV-{\it r}$>$=-1.5\,mag. 
Such approximation only provides a qualitative comparison,
however we believe is quite close to reality since the Galaxy model
includes WD in binaries in the predicted counts  but does not account for the color effects
of the companion stars.
The good qualitative  match between the observed surface density of 
objects and the Milky Way model, 
confirms our prediction (see above) that GALEX can detect
hot WDs 
throughout  the Galactic  halo. 
The GALEX selection of high-gravity hot stars in the Galaxy offers therefore
a major improvement with respect to previous surveys.  









For comparison to previous catalogs based on extensive surveys, we recall that
Acker et al. (1982) catalog of Galactic Planetary Nebulae 
includes 36 [12] objects at latitudes higher than 30 deg [45 deg],
out of a total of $\approx$1140 objects.
 McCook \& Sion (1999) catalog of WDs has 2364    [   1553]
WDs at latitudes higher than 30 deg [45 deg], out of $\approx$3060
total objects. Kleinmann et al.  (2004) catalogue 2551 certain 
white dwarf stars, 240 hot subdwarf
stars, and another 144 possible, but uncertain, white dwarf and hot
subdwarf stars from the 1360 deg$^2$ of SDSS DR1, or about 2.2 objects
per square degree. 
The Palomar-Green catalog of UV-excess stellar objects 
(Green et al. 1986), covering 10,714 square degrees,  
lists 1874 objects with limiting magnitudes between B=15.49 and
B=16.67. Eisenstein et al. (2006) catalogue about 9316 WDs
and 928 subdwarfs from the  SDSS Data Release 4, covering 4783 square degrees.
The GALEX selection of hot stellar objects is more sensitive to the hottest
objects, and especially to binaries containing a hot WD, since the 
optical photometry in such cases is dominated by the cooler companion
(see also Bianchi et al. 2006b, Figure 6, for an example).

The question of contamination of the hot-star candidate sample 
by non-stellar point-like  sources, 
 can be better investigated with follow-up spectroscopy, 
which is under way. 
In order to  assess the robustness of our photometric selection, 
we have matched our samples of photometrically selected 
 hot stars with the SDSS Data Release 5 (DR5) spectroscopic data.
The spectroscopic survey is less deep than the photometric
survey, so the fraction of our hot-star selected samples
that have available spectra in DR5 is higher for the AIS samples
than for the MIS samples. 
Of the hot stars (\Teff $>$ 18000~K) selected from the FUV-NUV color
(Figure  \ref{plot_LFNUV_HS}), 28\% (AIS) and 5\%(MIS) have optical  spectra. 
In particular, of the ``single'' hot-star candidates (whose optical colors
are consistent with the UV colors, and are plotted as solid-color
histograms in Figure  \ref{plot_LFNUV_HS}), 
29\%(AIS) and 16\%(MIS) have SDSS spectra.
Of our sample ``binaries'' (which  may
include both true stellar binaries with a WD plus 
cool companion and spurious objects),
   24\%(AIS) and 1\%(MIS) have  spectra.
Therefore, the statistics are more significant for AIS and for
``single'' WD selections than for other samples. 
Of the sources with spectroscopic classification, 97\%(AIS)/95\%(MIS) 
of our ``single'' hot-star candidates are confirmed as stars. 
Out of the 
``binary'' candidate sample, 
45\%(AIS) and 31\%(MIS) 
are spectroscopically classified as stars, 47\%(AIS) and 65\%(MIS)
are spectroscopically classified as QSOs. 
The fractions are very similar for the  sample of hot stars 
with \Teff $>$30,000K.
These numbers confirm our previous conclusions, based on  the  
photometrically histograms of density counts, 
that FUV-NUV selected hot stars with inconsistent optical colors
include both true stellar binaries 
and extra-galactic objects with SED differing
from the average QSO templates shown in our color-color diagrams. 
More importantly,  they indicate that our FUV-NUV selection 
of ``single'' hot stars is quite robust. 

 As for the hot stars (WD) selected from the \nggr diagram, 
and shown in the histograms of figure 8,
57\%(AIS)/40\%(MIS)  have SDSS spectra. Of these, 66\%(AIS)/61\%(MIS) 
 are spectroscopically confirmed as stars; 
30\%(AIS)/34\%(MIS) are spectroscopically classified as QSOs. 
We believe that the purity of the stellar sample is lower in this
case because the locus used to extract the high gravity stars includes
lower \Teff's than our FUV-NUV selection, and is close to the QSO locus.

\subsection{QSO Candidates}
\begin{figure}[!ht]
\centerline{
\vbox{\hbox{
\psfig{file=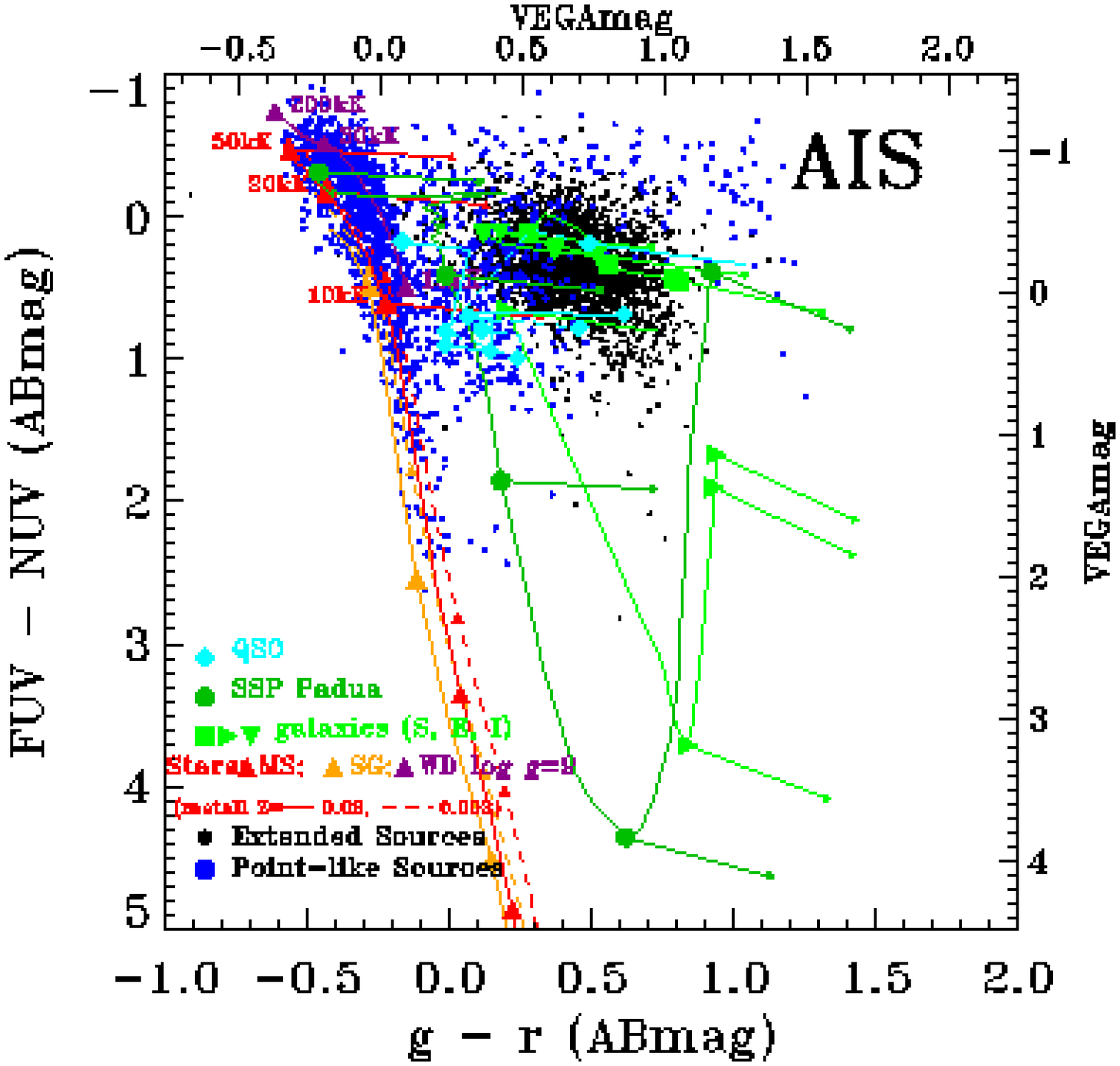,width=3.1in}
\psfig{file=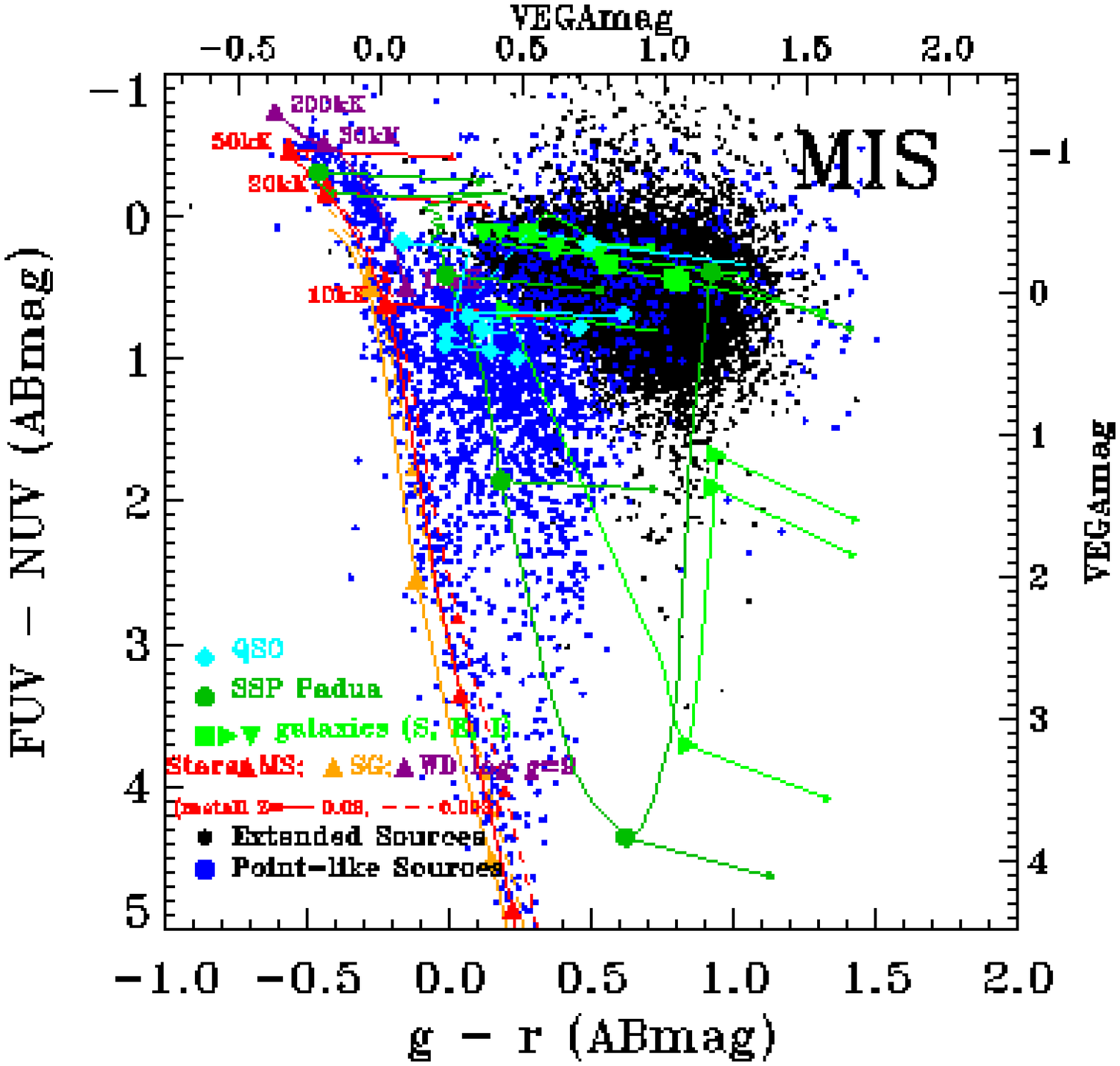,width=3.1in}
}
\hbox{
\psfig{file=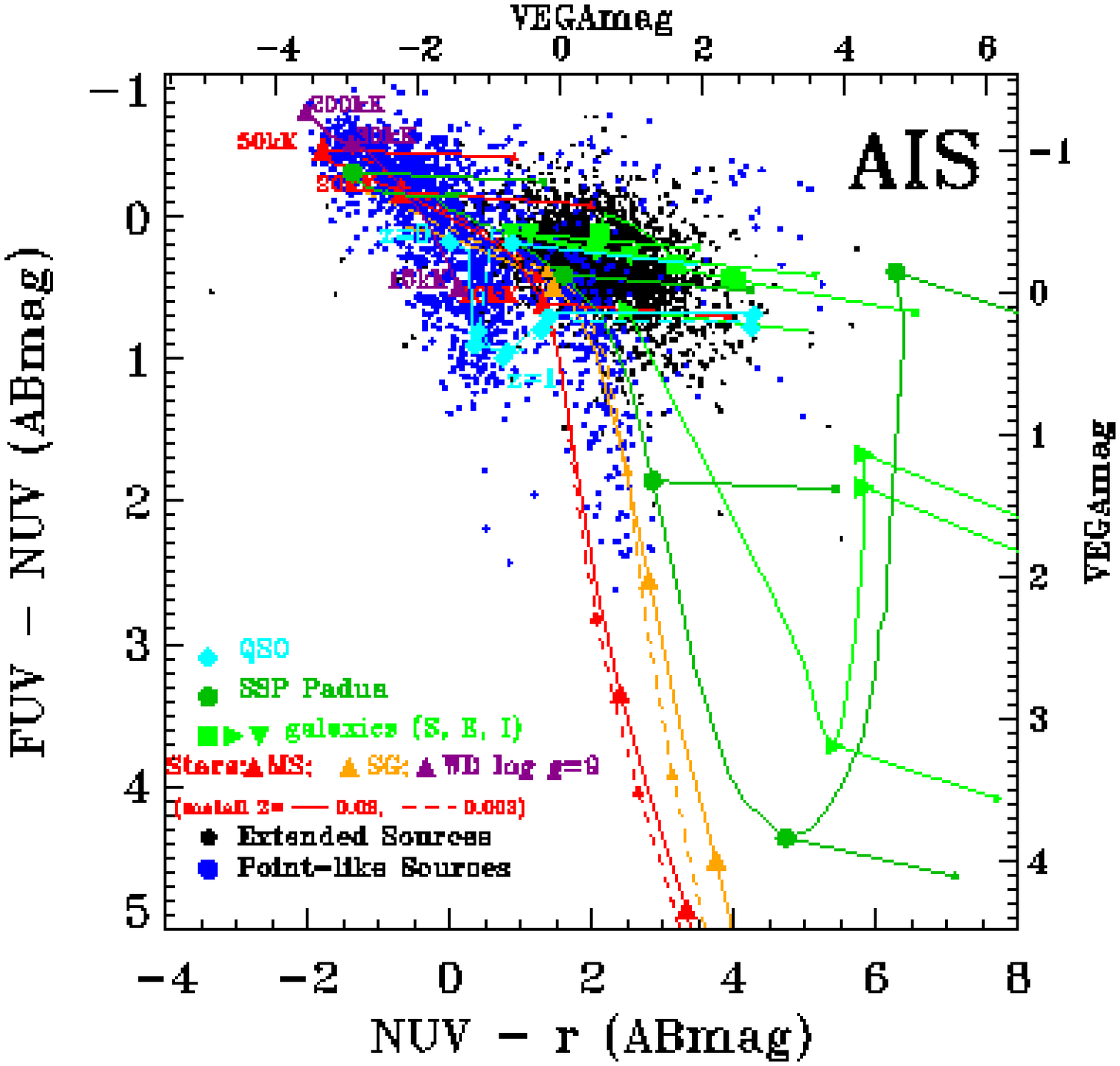,width=3.1in}
\psfig{file=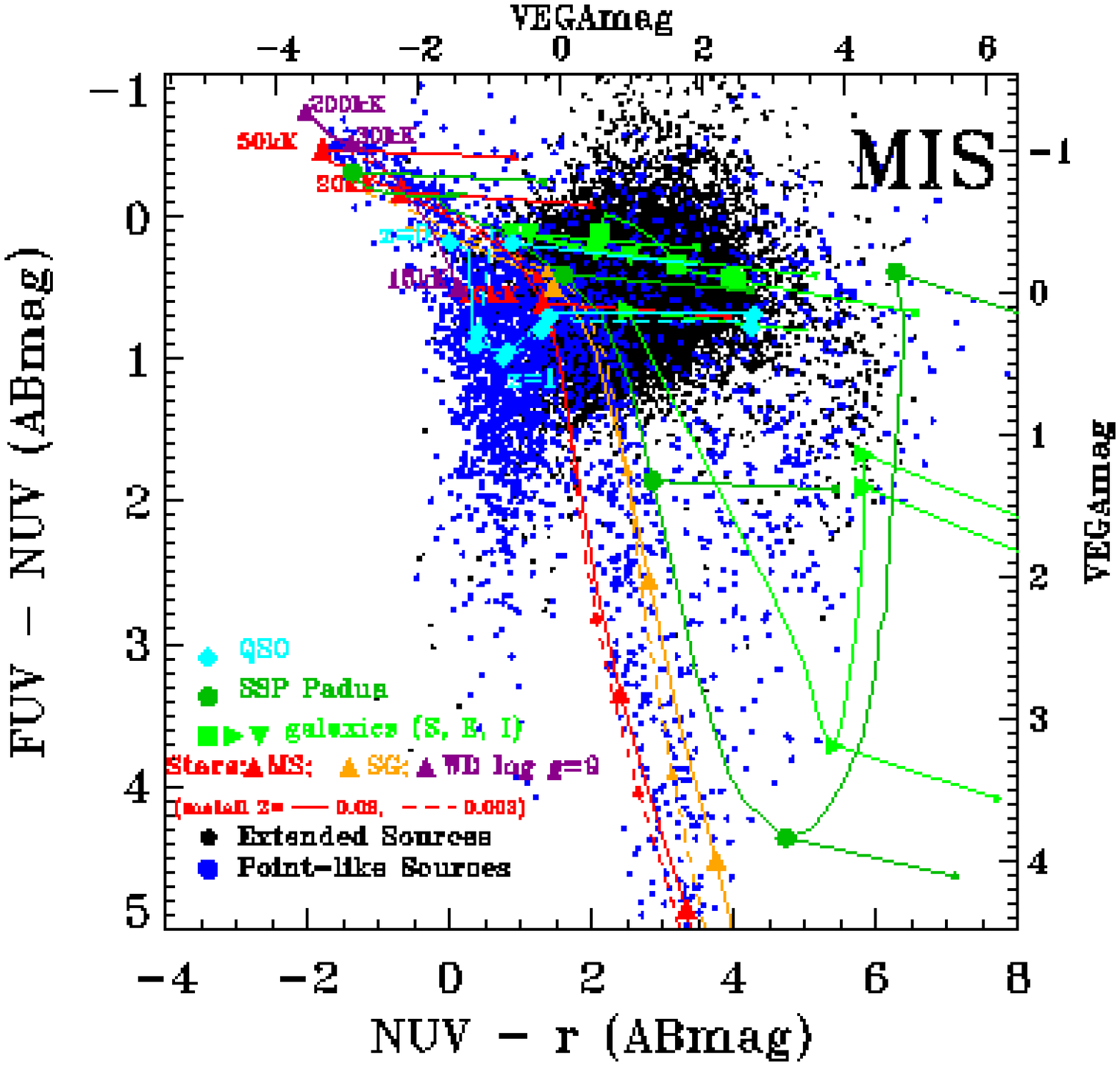,width=3.1in}
} } }
\caption{  \footnotesize The \fngr and \fnnr diagrams. 
Symbols and colors as in previous figure. 
  Stars and galaxies are well separated. The low-{\it z} QSOs 
are in between stars
and galaxies in the top diagrams but are more separated from both these
classes of objects in the \fnnr diagram,  especially for redshifts around 1
(the lowest point in the 'dip' of the QSO model colors). Here, 
QSOs at {\it z}=0 overlap with  cool WDs, and QSOs with 
{\it z}$\approx$ 1.6 have similar colors to A-type main-sequence stars, 
but the selection of QSOs around {\it z}=1 is best performed by using 
this color combination. 
Cataclysmic Variables at some stages 
(binaries containing a WD and an accretion disk)
may contaminate the
{\it z}=1 QSO locus. 
These objects are rare, and can be recognized
only  spectroscopically.
The \fnnr diagram can also be used to separate single and binary hot WDs
whose  FUV-NUV colors are bluer than the QSOs.
}
\label{plot_cc3}
\end{figure}

\begin{figure}[!ht]
\centerline{
\hbox{
\psfig{file=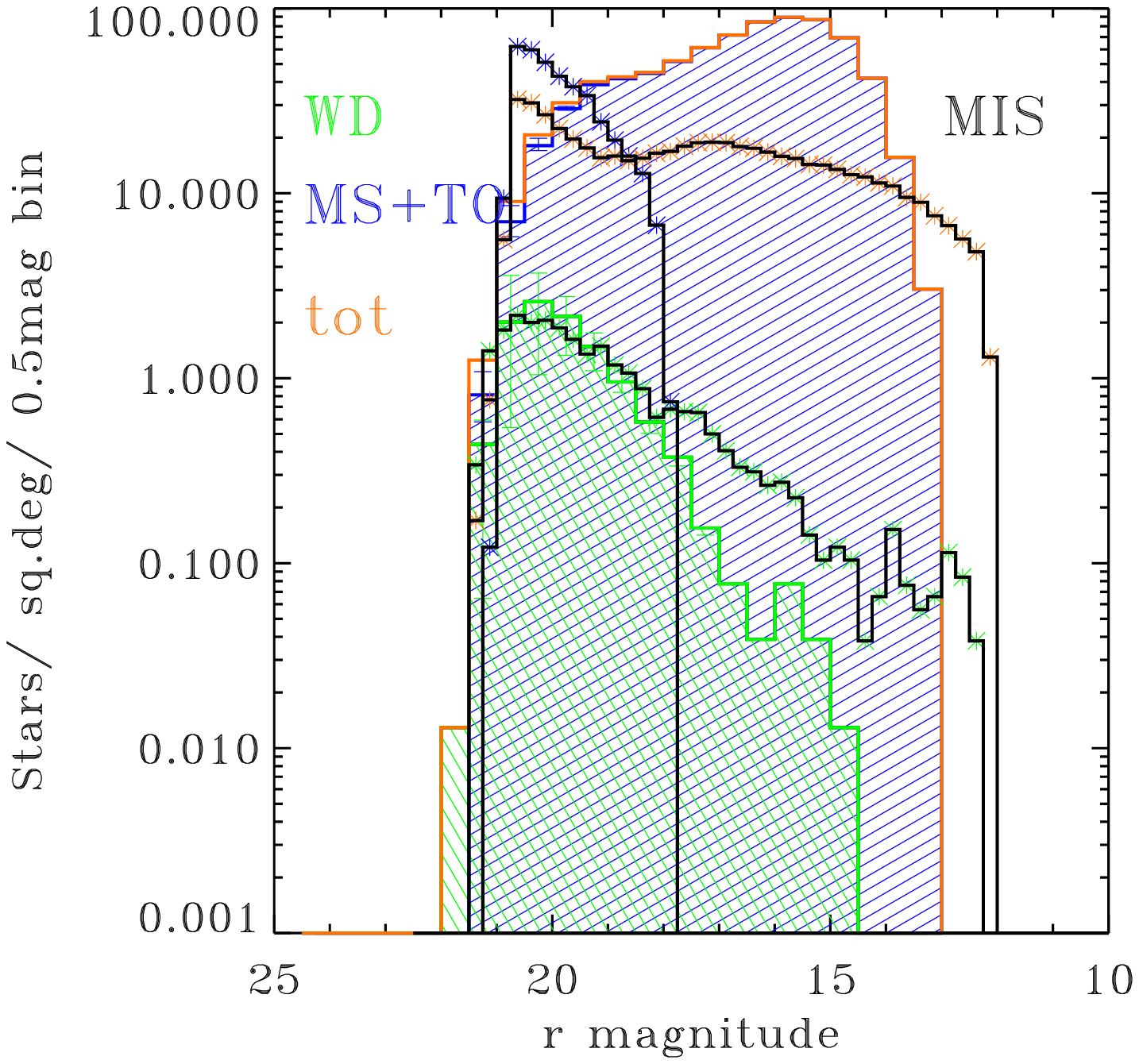,width=4.in}
\psfig{file=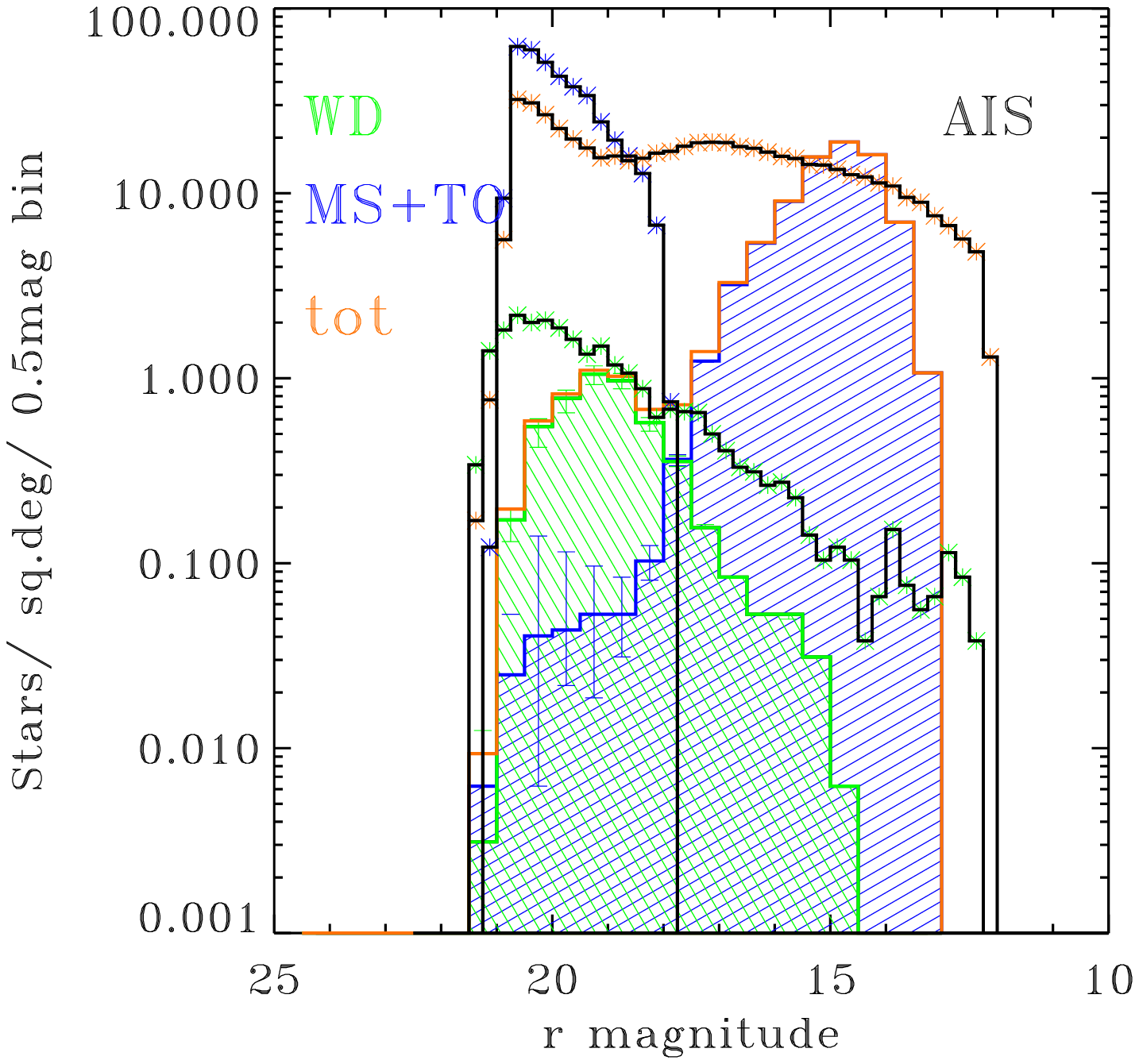,width=4.in}
} }
\caption{
The surface density of WDs (green) and lower gravity stars 
(blue).
 In orange, the sum of the two samples.
The stars were selected from the  \nggr colors, 
for direct comparison with the Milky Way model (see text). We restricted the sample
to sources with photometric errors better that 0.1mag in NUV, {\it g} and {\it r},
to avoid contamination by QSOs (see figure \ref{plot_cc1}).
 The diagram on the left shows the 
sources extracted from the deeper MIS survey, the one to the right shows 
the sources from the AIS.  The black histograms show the 
predictions from the Milky Way model (see text for details), with colored
asterisks to indicate  WD, non-WD, and total respectively.
Sources around 14th mag in optical bands may be affected by saturation in the
SDSS survey and therefore excluded, so our histograms are incomplete
for objects brighter than {\it r}$\approx$ 15mag. 
\label{plot_LF_WD} }
\end{figure}

\begin{figure}[!ht]
\centerline{
\hbox{
\psfig{file=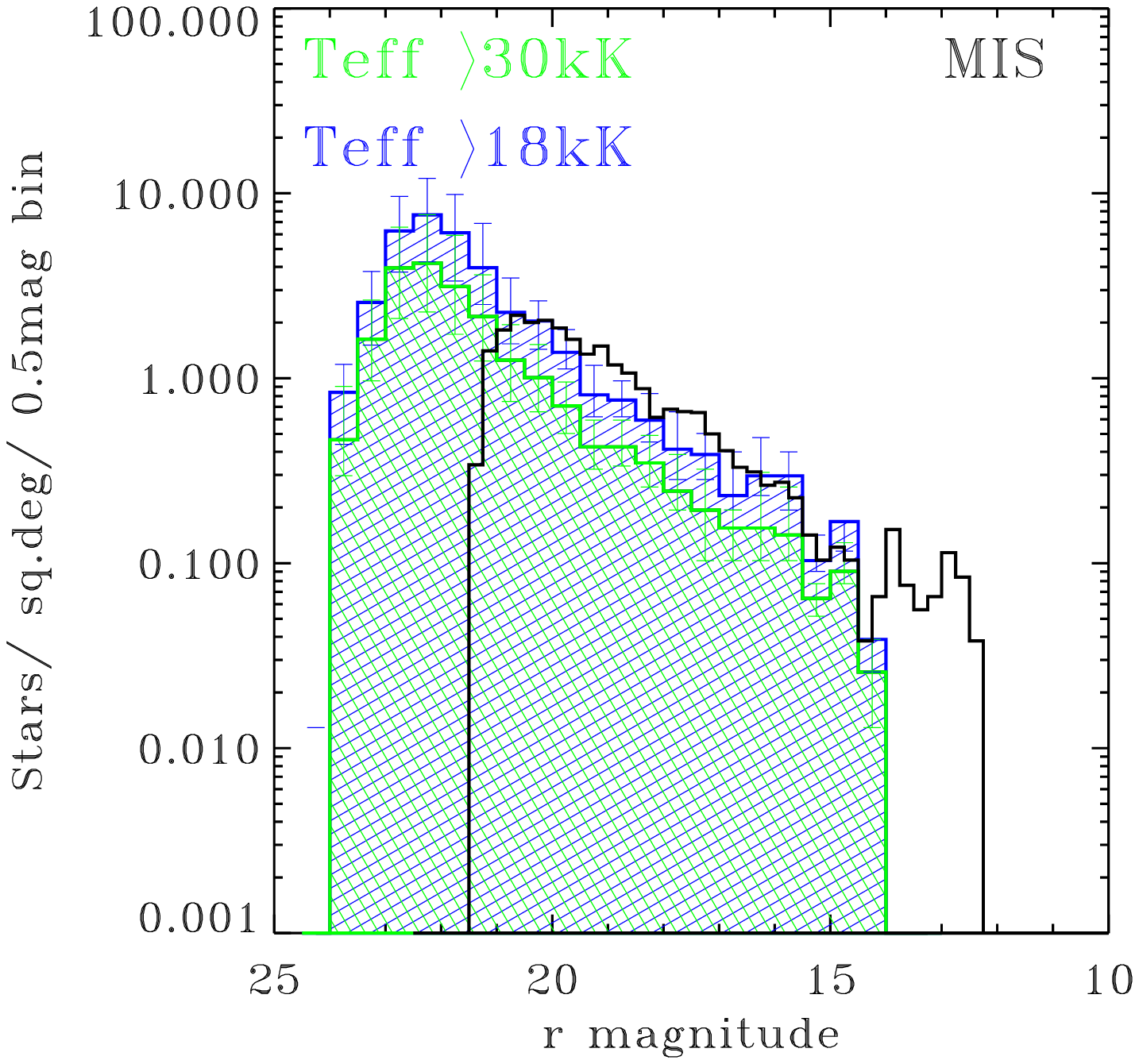,width=4.in}
\psfig{file=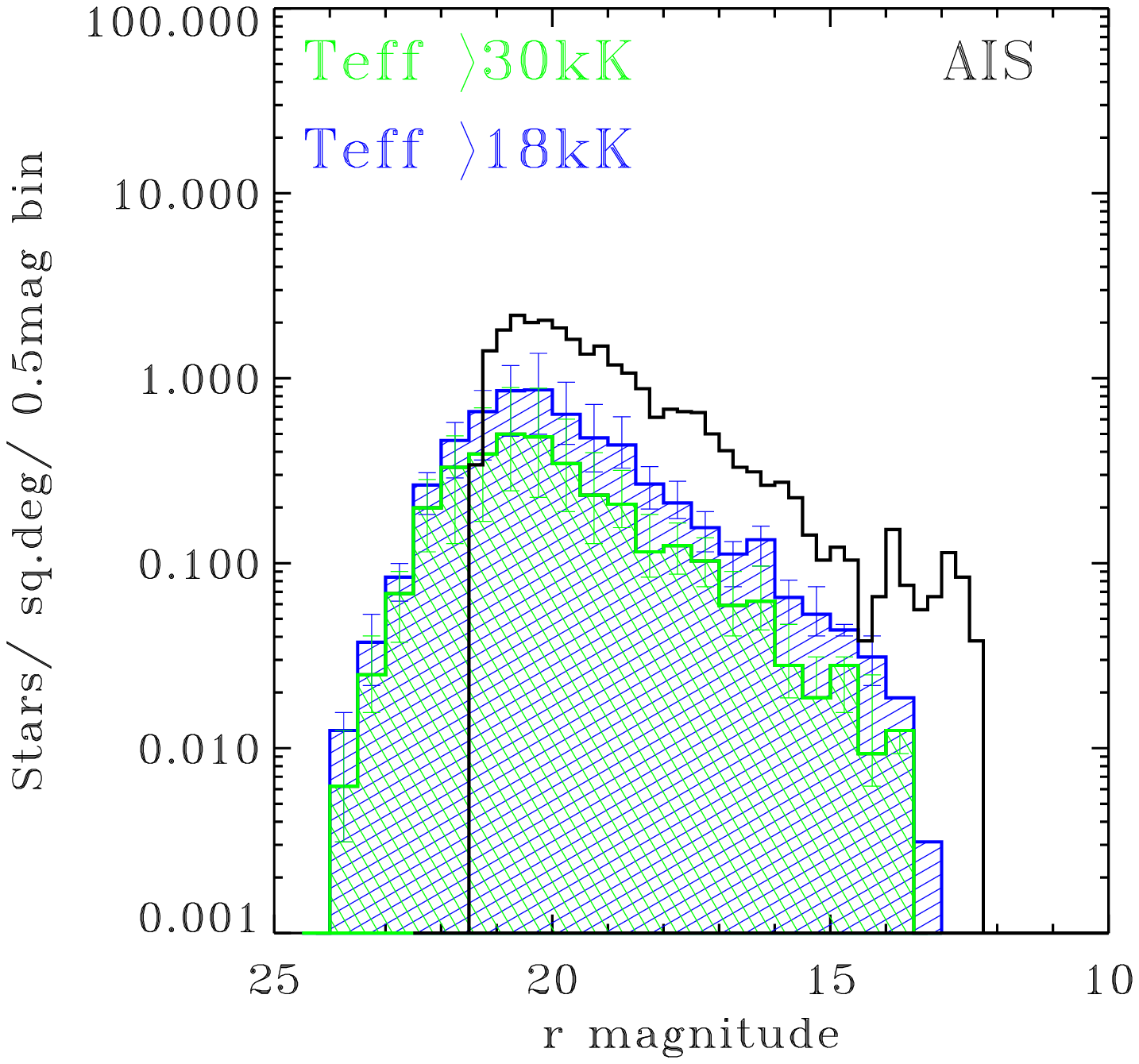,width=4.in}
} }
\caption{
The surface density
 (in the {\it r}-band) of hot stars selected from the FUV-NUV 
(GALEX) color. The FUV-NUV color cuts correspond to \Teff $>$ 18,000~K
(blue histogram) and  \Teff $>$ 30,000~K (green histogram) for stars with log~g=5.0,
but include lower \Teff' s for stars of higher gravities (see text).
Sources with photometric errors of $<0.3$ mag
in FUV and NUV are included. 
This selection does not use SDSS measurements, thus avoids the 
 loss of bright stars due to saturation seen in the previous figure.
  A fraction of these objects have optical colors discrepant from
the hot-WD UV colors, and they are either binaries with a cool companion or
extragalactic objects intruded in the sample. The apparent
difference in the AIS and MIS counts at bright magnitudes is
due to these objects having an {\it r} magnitude that
does not correspond to that of a single WD (which is not taken 
into account by the Milky Way model), as proven by
the next figure.   For the same reason, the apparent match
of the MIS number of objects 
 with the model is probably biased in the {\it r}-band. 
}
\label{plot_LF_HS} 
\end{figure}

\begin{figure}[!ht]
\centerline{
\hbox{
\psfig{file=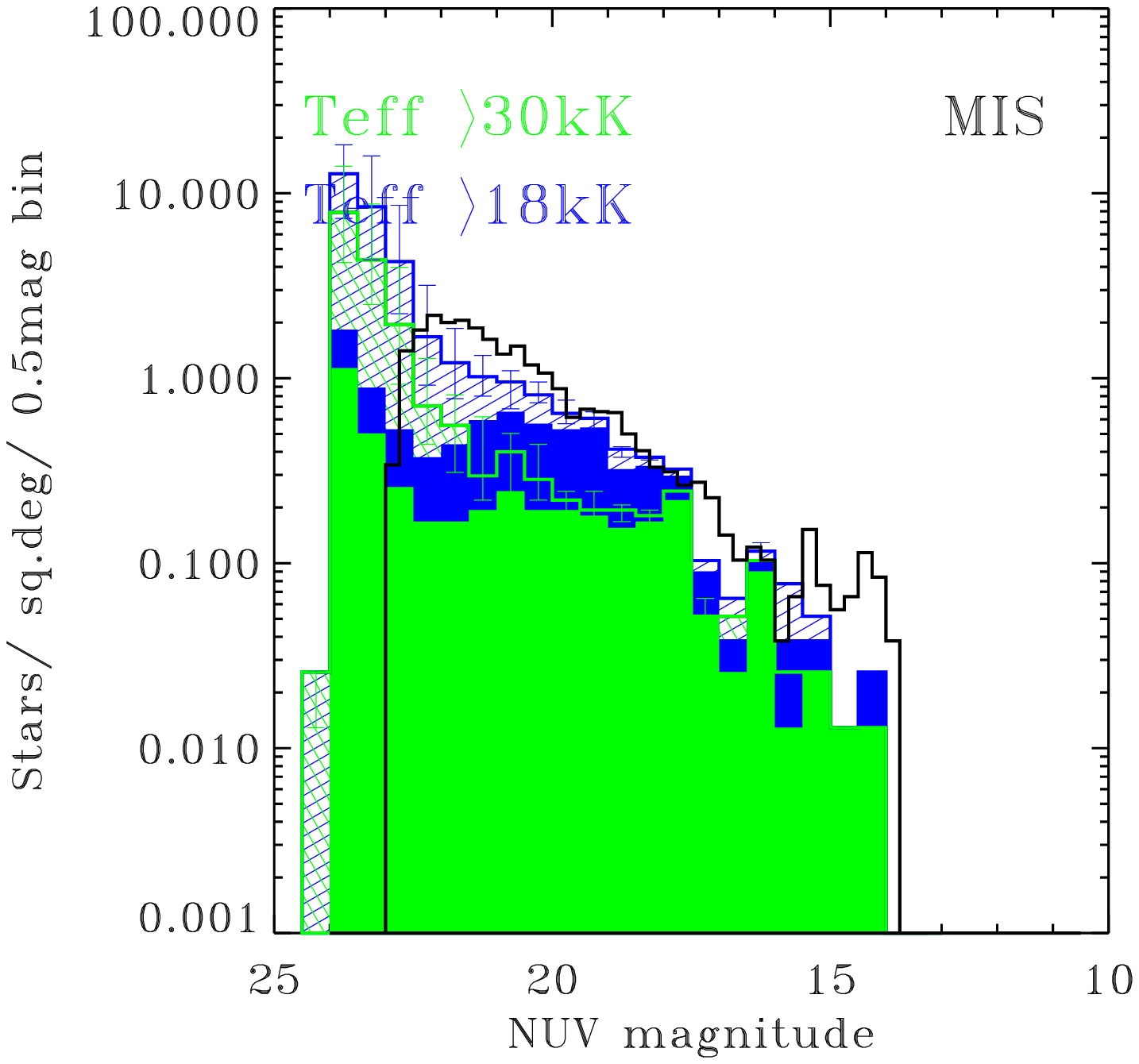,width=4.in}
\psfig{file=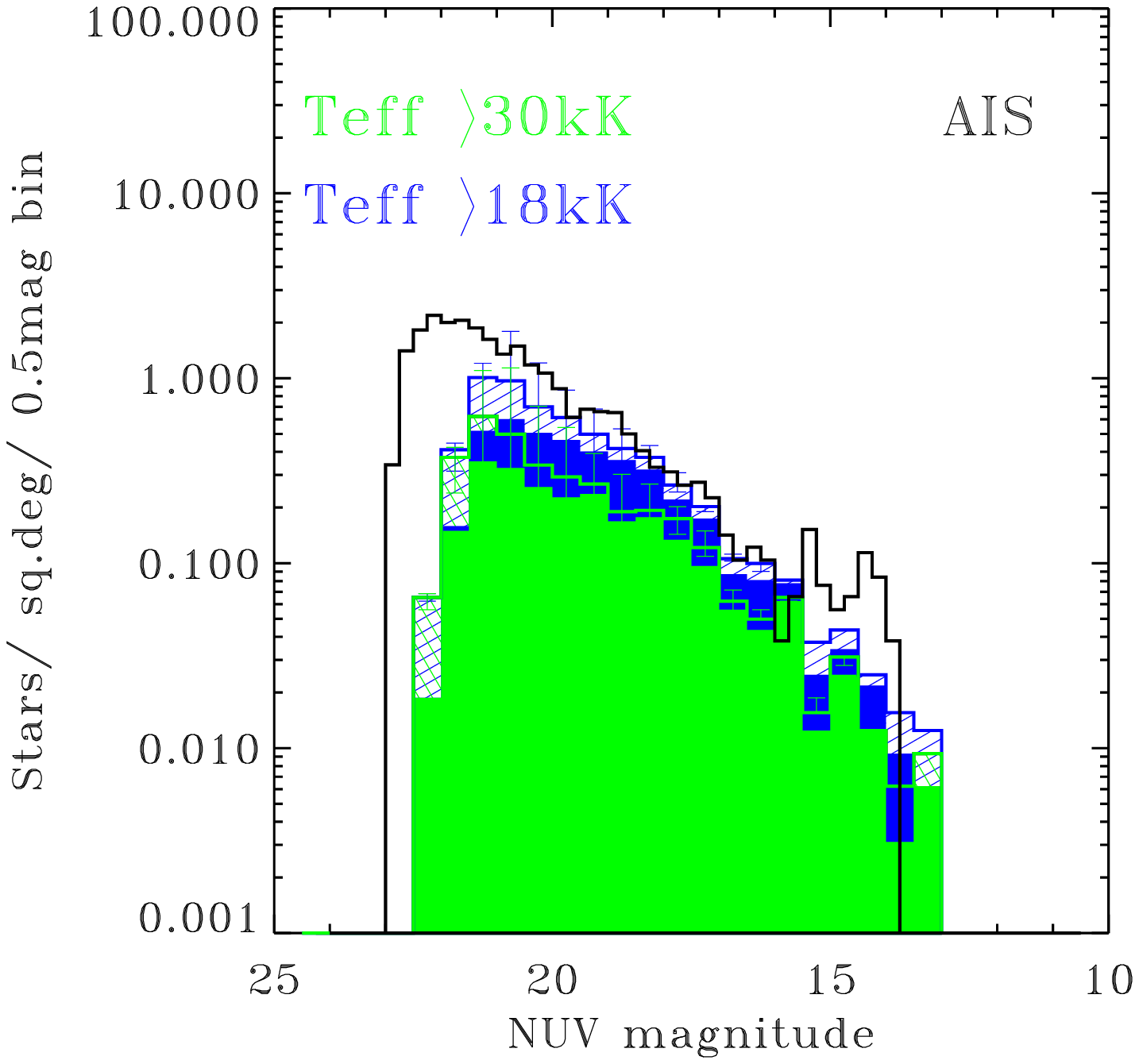,width=4.in}
} }
\caption{
The NUV-band surface density of the hot stars selected from FUV-NUV 
GALEX color, same sample as in the previous Figure.
Solid-color histograms show ``single'' hot WD-candidates, dashed
 histograms include both single-WD  and objects with optical colors
redder than what correponds to a hot-WD with the observed UV color. These include
binaries with a hot-WD and  possibly also 
extragalactic objects 
as suggested by their number increasing
at fainter magnitudes. The Galaxy {\it r}-band WD model (black line)
 has been shifted using 
an average color of $<$NUV-r$>$=-1.5~mag.  
Because the UV band is more representative of the WD component
 in the case of binaries, the AIS and MIS counts at bright magnitudes
are consistent as we expect, until the AIS becomes
incomplete. The AIS is better sampled because it covers a larger area. 
}
\label{plot_LFNUV_HS} 
\end{figure}

\begin{figure}[!ht]
\centerline{
\psfig{file=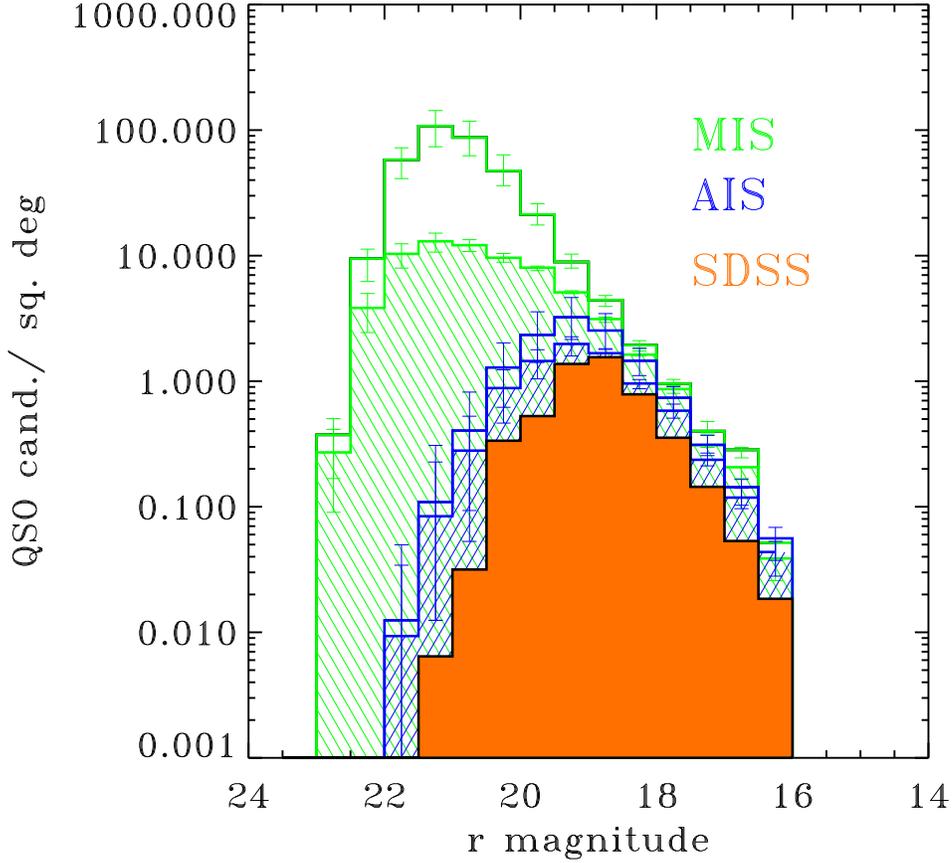,width=5.in}
}
\caption{
The surface density of low-redshift QSO candidates from the 
GALEX catalogs. 
In green,  
sources extracted from the deeper MIS, in blue the QSO candidates
from AIS.  Histograms only outlined 
include both point-like and extended sources in the color locus of 
the QSOs. Shadow-filled histograms 
include only point-like sources with the same color selection. 
Error bars are shown. 
 The solid orange histogram shows the 
SDSS QSO catalog of 
Schneider et al. (2005);  
included are QSOs with   ({\it z}$<$1.3), to match the GALEX color selection.
Including QSOs at all redshifts from the SDSS catalog does not change significantly
the SDSS histogram on our scale.
Note that the SDSS QSOs are spectroscopically confirmed;
they are about 60\% to 85\% of the SDSS QSOs candidates selected photometrically.
We estimate that our photometric selection may contain 15\% spurious sources. 
 \label{plot_LF_QSO} }
\end{figure}

As  pointed out previously by Bianchi et al. (2005), 
the \fnnr diagram  (Figure \ref{plot_cc3}) can also be used 
to select low redshift QSOs candidates. Figure \ref{plot_LF_QSO} shows
the observed surface density of low-redshift QSO candidates, 
selected from this color combination from both the MIS and AIS surveys.
For comparison, the recent QSO catalog of Schneider et al. (2005) 
 from the SDSS DR3-release is also shown (orange histogram). 
To normalize the SDSS QSO catalog to density of objects per 
 unit area, we used an area of 3732 
square degrees, the coverage of the spectroscopic DR3 release,
because the catalog includes spectroscopically confirmed QSOs. 
We also limited the SDSS catalog to QSOs with {\it z}$<$1.3, corresponding
to our GALEX photometric selection.
To actually scale the SDSS spectroscopically confirmed catalog to 
the SDSS photometric selection of QSO candidates,  we must consider
that about 60\% of the photometrically selected candidates have
been observed spectroscopically so far (Richards, private comm.)
although the exact fraction is not available from the Schneider et al. (2005) 
catalog.  For more discussion, see Richards et al. (2005) 
and Hutchings et al. (in preparation). The density of objects 
in our GALEX photometrically selected sample is larger than the SDSS
QSO candidate sample, and  extends to fainter
magnitudes. 

 In Figure  \ref{plot_LF_QSO}, we show separately the point-like 
QSO-candidate GALEX sources,
and all sources with QSO-like colors regardless of
spatial extent, i.e. including extended sources which have observed
colors consistent with our photometric selection.
The present classification of ``point-like'' and ``extended'' is based
on the SDSS $\approx$1.4$''$ psf, and obtained from the pipeline.
However, the contrast
between the AGN central source and the underlying galaxy is still
poorly characterized at low redshifts ({\it z}$\approx$1). The GALEX-selected
catalog of low redshift QSO candidates 
presents an opportunity to clarify this issue, which
 will be investigated with follow-up deeper imaging.

We finally point out  that other objects share the locus of the QSOs in 
the color-color diagram: cataclysmic variables (CV)
with a significant accretion disk have  similar 
colors and thus potentially contaminate the QSO sample. 
These objects however are extremely rare, 
about 0.02 per square degrees (e.g. Szkody et al. (2004) 
 and references therein). 
Our QSO candidate density is at least over one order of magnitude 
higher than the expected density of CV in the AIS, and about two dex higher in the MIS, 
therefore we expect 
our GALEX-selected QSO candidates to not be significantly contaminated.
However, such stellar sources (expected to be nearby and therefore bright)
might explain the excess of sources at bright magnitudes,
with respect to the confirmed SDSS QSOs.
To  estimate the purity of the photometrically-selected sample,
as we did for the hot stars samples, we have matched our QSO candidates 
with the SDSS DR5 to search for archival optical spectra. 
We found that 60\%(AIS)/17\%(MIS) of our QSO candidates (point-like)
 have spectra, and of these,
 83\%(AIS)/85\%(MIS) 
confirm the QSO classification. 
The statistics is fairly significant: 1631 objects have spectra 
 out of 2689 total objects for AIS, 
930 out of 5312 for MIS.  Of the non-QSO objects, 14\%(AIS)/12\%(MIS) 
are spectroscopically classified as stars, consistent with what one
can qualitatively expect from the color-color diagrams.  Therefore, 
 based on the results from  the SDSS automated spectral classification, the
purity of our GALEX-selected point-like QSO-candidate sample can be
considered $\approx$ 85\%,  comparable or higher to the 
success rate of optical selection. 
As for the extended sources that fall in our QSO-selection color locus, 
 only 19\%(AIS)/$<$0.1\%(MIS)    
have spectra:  of these, 27\%(AIS)/48\%(MIS) are spectroscopically
 classified as QSOs.
 The statistic is less significant than for the point-like sources.

\acknowledgments

GALEX (Galaxy Evolution Explorer) is a NASA Small Explorer, launched in April 2003.
We gratefully acknowledge NASA's support for construction, operation,
and science analysis of the GALEX mission,
developed in cooperation with the Centre National d'Etudes Spatiales
of France and the Korean Ministry of 
Science and Technology.  
We are grateful to John Hutchings,  Wei Zheng and 
Gordon Richards for discussions about QSO issues and 
clarifications about the QSO SDSS catalogs and templates, 
to Alessandro Bressan for
providing the yet unpublished SSP models and for extremely useful discussions,
and (with Olga Vega) for assistance in calculations of the galaxy template,
to Paula Szkody and Knox Long for illuminating discussions about CVs 
and for the  CV templates.
{\it Facilities:} \facility{GALEX}, \facility{Sloan}

~ \vskip -2.cm

\clearpage
{\footnotesize
\begin{deluxetable}{lcccccc}
\tablewidth{0pt}
\tablecaption{Statistics of sources in the different bands {\label{tlimits}} }
\tablehead{
\multicolumn{1}{l}{ }                  & \multicolumn{6}{c}{Number of Sources}  \\  
\multicolumn{1}{l}{Band} & \multicolumn{2}{c}{(error$<$0.2 mag)} & \multicolumn{2}{c}{(error$<$0.1 mag)}  &  \multicolumn{2}{c}{(error$<$0.05 mag)} \\
\multicolumn{1}{l}{ } & \multicolumn{1}{c}{AIS} & \multicolumn{1}{c}{MIS} & \multicolumn{1}{c}{AIS} & \multicolumn{1}{c}{MIS} &  \multicolumn{1}{c}{AIS} & \multicolumn{1}{c}{MIS} \\
}
\startdata
\fuvmag    &  15007 & 137940 &   2902 &  23840 &    821 &   3866 \\
\nuvmag    & 250063 & 627353 &  83540 & 314498 &  31275 &  75680 \\
{\it u}    & 269070 &  99563 & 206481 &  74777 & 169439 &  58671 \\
{\it g}    & 631119 & 287428 & 448334 & 172208 & 312059 & 116371 \\
{\it r}    & 699744 & 355341 & 513223 & 211791 & 350779 & 132569 \\
{\it i}    & 637288 & 310262 & 448761 & 182945 & 309971 & 117935 \\
{\it z}    & 365375 & 144429 & 255060 &  96299 & 194863 &  73006 \\
\enddata
\label{tbl-1}
\end{deluxetable}
}
\normalsize

\end{document}